\documentclass[aps,preprint]{revtex4}
\usepackage{amsfonts}
\usepackage{amsmath}
\usepackage{amssymb}
\usepackage{graphicx}
\usepackage{subfigure}
\usepackage{mathrsfs, mathtools}
\usepackage{pdfpages}
\usepackage{multirow}
\usepackage{float}
\usepackage{color}
\usepackage{xcolor}
\usepackage{orcidlink}
\usepackage{caption}
\usepackage{lineno,hyperref}
\hypersetup{colorlinks =true, allcolors = blue}
\providecommand{\U}[1]{\protect\rule{.1in}{.1in}}

\begin{document}
\begin{center}
{\Large Astrophysical properties of static black holes embedded in a Dehnen type dark
matter halo with the presence of quintessential field}

Ahmad Al-Badawi \orcidlink{0000-0002-3127-3453}\\
Department of Physics, Al-Hussein Bin Talal University, P. O. Box: 20, 71111,
Ma'an, Jordan. 
\bigskip E-mail: ahmadbadawi@ahu.edu.jo\\
Sanjar Shaymatov \orcidlink{0000-0002-5229-7657}

Institute for Theoretical Physics and Cosmology, Zhejiang University of Technology, Hangzhou 310023, China\\
Institute of Fundamental and Applied Research, National Research University TIIAME, Kori Niyoziy 39, Tashkent 100000, Uzbekistan\\
University of Tashkent for Applied Sciences, Str. Gavhar 1, Tashkent 100149, Uzbekistan\\
Western Caspian University, Baku AZ1001, Azerbaijan\\
\bigskip E-mail: sanjar@astrin.uz\\

{\Large Abstract} \end{center}

From an astrophysical perspective, the composition of black holes (BHs), dark matter (DM), and dark energy can be an intriguing physical system. In this study, we consider Schwarzschild BHs embedded in a Dehnen-type DM halo exhibiting a quintessential field. This study examines the horizons, shadows, deflection angle, and quasinormal modes (QNMs) of the effective BH spacetime and how they are affected by the dark sector. The Schwarzschild BH embodied in a Dehnen-type DM halo exhibiting a quintessential field possesses two horizons: the event horizon and the cosmological horizon. We demonstrate that all dark sector parameters increase the event horizon while decreasing the cosmological horizon. We analyze the BH shadow and emphasize the impact of DM and quintessence parameters on them. We show that the dark sector casts larger shadows than a Schwarzschild BH in a vacuum. Further, we delve into the weak gravitational lensing deflection angle using the Gauss-Bonnet theorem (GBT). We then investigate QNMs using the 6th order WKB approach. To visually underscore the dark sector parameters, we present figures that illustrate the impact of varying the parameters of the Dehnen-type DM halo as well as quintessence. Our findings show that the gravitational waves emitted from BHs with a dark sector have a lower frequency and decay rate compared to those emitted from BHs in a vacuum.  

\section{Introduction}

It is widely believed that general relativity (GR) has been regarded as the most successful and tested theory of gravity. Yet, it has notable challenges in explaining fundamental questions, such as the occurrence of singularities inside black holes (BHs), where GR loses its applicability \cite{Borde03PRL,Borde94PRL}. Additionally, GR is still not capable of explaining phenomena such as dark matter (DM) and dark energy.
These challenges indicate that GR is an incomplete theory and needs to be modified. From this perspective, the issues faced by GR might be resolved by proposing modified gravity theories and exploring their unique aspects as potential explanations for these issues in an astrophysical scenario. It is worth noting, however, that recent modern observations (i.e., gravitational waves (GWs) \cite{Abbott16a,Abbott16b} and the BH shadow \cite{Akiyama19L1,Akiyama19L6}) will be crucial in providing accurate and promising insights into the unique features of these GR issues. Taken altogether, it is important to explore the nature of existing gravitational fields in the context of alternative theories of gravity to provide a comprehensive explanation for these issues. 

From an astrophysical view point, BHs cannot be found in a vacuum due to the presence of surrounding matter fields in their nearby environment. For example, observations of supernova explosions (SNIa) have confirmed the accelerated expansion of the universe through a cosmological constant $\Lambda$ acting as vacuum energy and a repulsive gravitational effect. In this regard, the effect of the cosmological constant plays a pivotal role at large scales and in the BH surrounding environment, causing the universe to accelerate because of the vacuum energy $\Lambda$, which is interpreted as dark energy in the Einstein equation. It is to be emphasized that the cosmological constant can be estimated to be of the order of $\Lambda \sim 10^{-52}m^{-2}$. However, it is widely believed that the repulsive effect of the cosmological constant has a significant impact on the accelerating rate of expansion of the universe at large scales~\cite{Peebles03,Spergel07}, and its impact has since been extensively investigated in various astrophysical contexts \cite{Stuchlik05,Cruz05,Stuchlik11,Grenon10,Rezzolla03a,Arraut15,Faraoni15,Shaymatov18a}. Additionally, it is important to note that there are alternatives to explain the behavior of dark energy. For example, the quintessence field, proposed by Kiselev \cite{Kiselev03}, has also been considered an alternative model to explain the behavior of dark energy~\cite{Peebles03,Wetterich88,Caldwell09}. In this solution, the  quintessence field is defined by the equation of state $p = \omega_{q}\rho$ with $\omega_q\in (-1;-1/3)$ \cite{Kiselev03} and $(-1;-2/3)$ \cite{Hellerman2001JHEP}.

It is a well-established fact that the motion of massive particles can be strongly affected by the geometry in the strong field regime, while their geodesics, together with the background geometry, can be altered by matter fields existing in the BH surrounding environment. From this viewpoint, dark matter (DM) would come into play in explaining such phenomena, leading to the introduction of DM through the observation of the flat rotation curves of giant spiral and elliptical galaxies. Later, relying on astrophysical data, it was found that the elusive DM can only explain the rotational velocity of stars orbiting giant spiral galaxies, resulting in DM contributing to nearly 90 \% of the entire mass of the galaxy, with approximately 10 \% luminous matter formed by baryonic matter \cite{Persic96}. In the early evolution of the universe, stars were formed with contribution of in regions, especially very close to the galaxy centers. Interestingly, the observations later revealed that, due to various dynamical scenarios, DM gradually drifted outward to form a galactic dark matter halo in the surrounding regions of the host galaxies. Therefore, based on the galaxy formation scenario, in most cases all giant elliptical and spiral galaxies are hosted by a supermassive BH at the galactic center surrounded by a DM halo (see, e.g., \cite{Akiyama19L1,Akiyama19L6}). Various black hole solutions incorporating a DM profile have been proposed in the context of DM scenarios. For example, inspired by a quintessential scalar field developed by Kiselev \cite{Kiselev03}, Li and Yang \cite{Li-Yang12} presented a BH solution with a DM distribution using a phantom scalar field without a cosmological role. In this line, there has been an extensive analysis in various situations since then (see, e.g., \cite{Hendi20,Rizwan19,Narzilloev20b,Shaymatov21d,RayimbaevShaymatov21a,Shaymatov21pdu,Shaymatov22prd}). There have also been models developed for a DM halo solution (see, e.g., \cite{Cardoso22DM,Shen24PLB,Hou18-dm}). Interestingly, a new DM halo solution has recently been proposed using a Dehnen-(1, 4, 0) type DM halo density profile \cite{Dehnen93}, in which the spherically symmetric BH spacetime is considered to be immersed in a DM halo profile \cite{dn14}. It should be noted that the phenomenological Dehnen DM density profile has been extensively investigated for various scenarios including galaxies \cite{Dehnen93}. In this paper, following \cite{Kiselev03,dn14}, we consider a spherically symmetric BH with a Dehnen-(1, 4, 0) type DM halo together with a quintessence field and explore it, along with its remarkable aspects and features, using BH observational properties in an astrophysical context such as shadow, deflection angle and quasinormal modes (QNMs).

The EHT obtained a photograph of the BH in the Milky Way galaxy's Sagittarius A* (SgrA*) and discovered that the radius of the ring around SgrA* is within $10\%$ of GR's prediction \cite{qqn1,qqn2,qqn3,qqn4}. This image clearly shows a bright ring-like structure outside the BH's shadow, which corresponds to the light emitted by the BH's accretion disk. The BH shadow 
appears as a dark region surrounded by a ring of light in the sky. Therefore, 
studying BH accretion disks can help us understand
characteristics of BHs and the impact of the dark sector, which includes Dehnen-type DM and the quintessence field. This dark sector/disk contributes to the formation of a BH shadow by trapping light that cannot escape due to strong deflection \cite{Synge66,Luminet79}. Analyzing the BH shadow and the gravitational lensing effects are crucial for exploring the geometry near the BH horizon. Numerous studies have extensively investigated BH shadow analysis in a variety of contexts \cite[see, e.g.,][]{Amarilla13,Konoplya19,Vagnozzi19,Afrin21a,Atamurotov16EPJC,Konoplya19PRD,Atamurotov21JCAP,Mustafa22CPC,Tsukamoto18,Asukula24,Rosa23b,Moffat20,Al-BadawiCTP24,Al-BadawiCPC24,Hendi23,newa1}.  Additionally, the gravitational lensing effects initially caused GR to be the most successfully tested theory of gravity to explain the unique aspects of BH geometry \cite{Eddington1919GL}. Therefore, testing the BH geometry is still an important task of GR. There has also been an extensive analysis done regarding the gravitational lensing around BHs in various gravity models~\cite[see, e.g.,][]{Bisnovatyi-Kogan2010a,Tsupko12,Cunha20a,Babar21a,Javed22GRG,Jafarzade21a,Atamurotov22,Atamurotov21galaxy,Kumaran_2023,Mizuno_2018,Rahvar19MNRAS,Izmailov19MNRAS}.

In addition, quasinormal modes (QNMs) are important characteristics of a BH's late-time response to external perturbations. Gravitational interferometers are used to detect quasinormal waves caused by spacetime perturbations \cite{Abbott16a,Abbott16b,Akiyama19L1}. Since the discovery of QNMs, the study of BH perturbations has become increasingly important in the last decade \cite{newa2,newa3}. QNMs dominate the late-time gravitational wave signals of BHs and other compact objects. QNMs are described by complex frequencies, with the real component corresponding to the oscillation frequency and the imaginary part relating to the damping rate caused by radiation emission. QNMs are also useful for determining the stability of black holes. In the case of positive imaginary parts of the quasinormal frequencies, the BH becomes unstable to perturbations, while in the case of negative imaginary parts, it remains stable.
 
The paper is structured as follows: Sec.~\ref{sec2} provides a description of the Schwarzschild BH embedded in a Dehnen-type DM halo in the background of quintessence field. Sec.~\ref{sec3} analyzes BH shadows and the impact of the BH parameters. In Sec.~\ref{sec4},  we investigate the deflection angle by adapting the method of the Gauss-Bonnet theorem. Next, in Sec.~\ref{sec5}, we evaluate and examine the QNMs. Finally, our conclusions are presented in Sec.~\ref{sec6}.

\section{Review of the BH metric geometry} \label{sec2}
\begin{figure}
    \centering
    \includegraphics[width=0.5\linewidth]{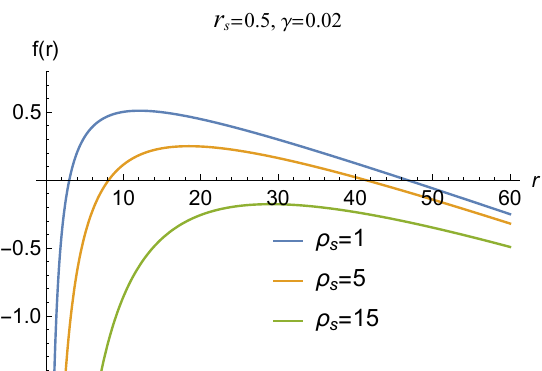}
    \caption{Metric function is plotted for different values of the BH parameters. Here, $M=1$ and $\epsilon=-2/3$.}
    \label{lapse}
\end{figure}

\begin{figure}
\begin{center}
\includegraphics[scale=0.4]{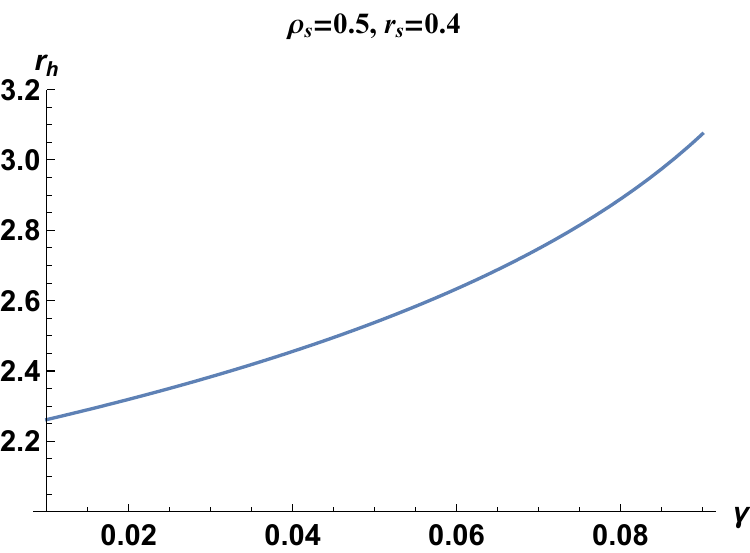}
\includegraphics[scale=0.4]{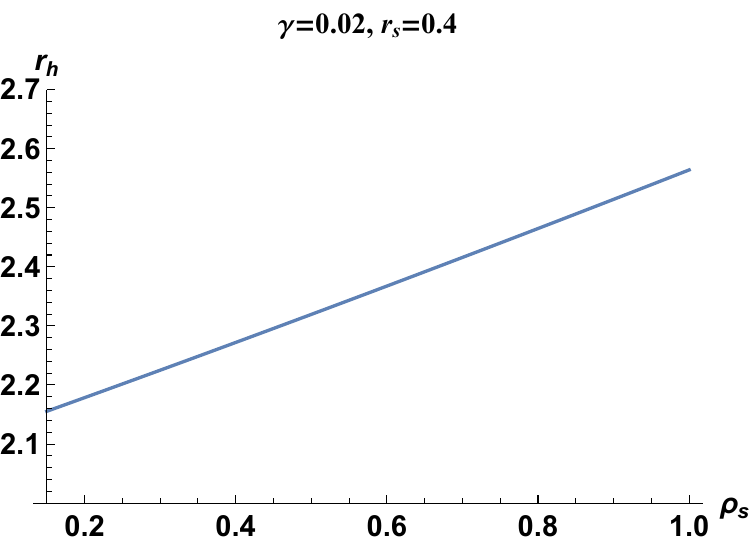}
\includegraphics[scale=0.4]{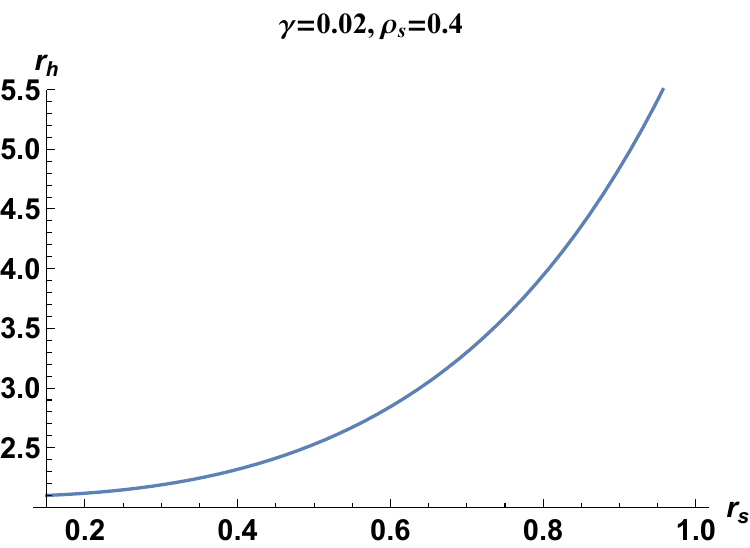}
\end{center}
\caption{Variation of event horizon $r_h$ with $\gamma$ keeping $\rho_s$ and $r_s$ (left) with $\rho_s$ keeping $\gamma$ and $r_s$ (middle) with $r_s$ keeping $\gamma$ and $\rho_s$ (right). }\label{hor12}
\end{figure}

\begin{figure}
\begin{center}
\includegraphics[scale=0.4]{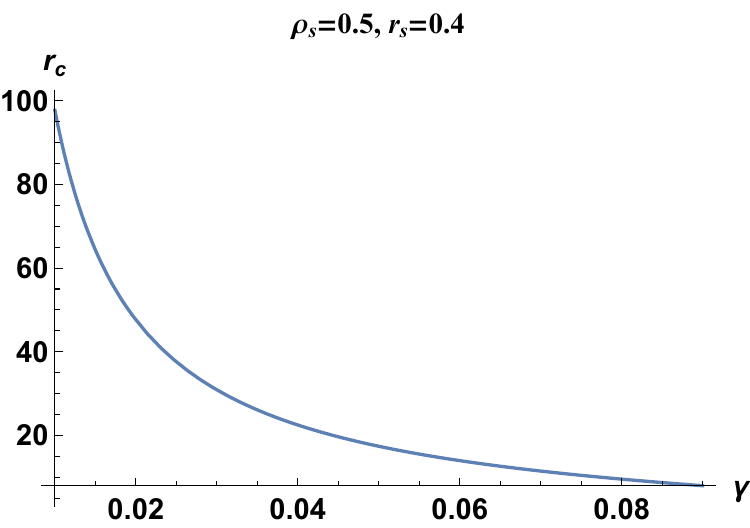}
\includegraphics[scale=0.4]{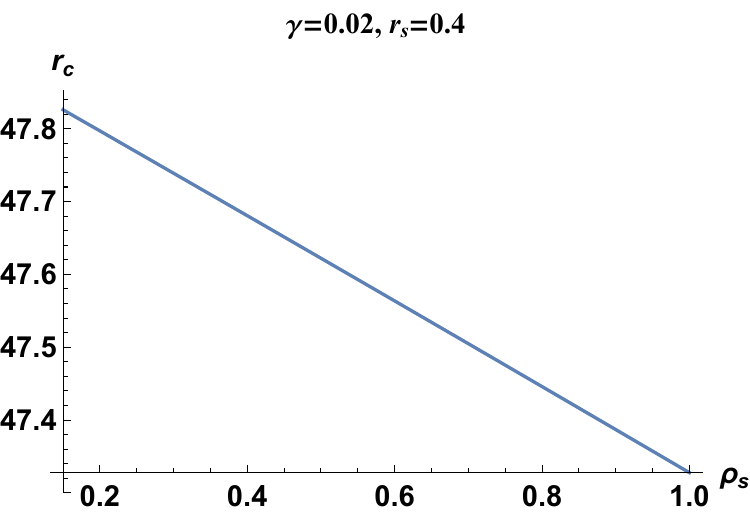}
\includegraphics[scale=0.4]{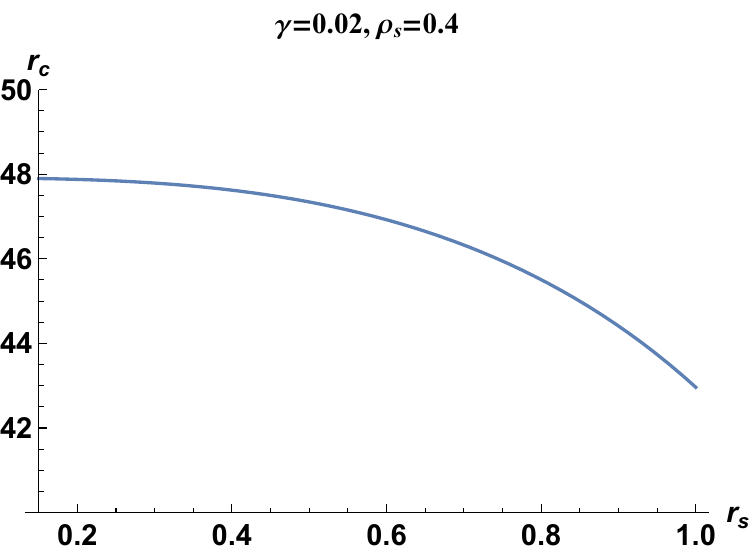}
\end{center}
\caption{Variation of cosmological horizon $r_c$ with $\gamma$ keeping $\rho_s$ and $r_s$ (left) with $\rho_s$ keeping $\gamma$ and $r_s$ (middle) with $r_s$ keeping $\gamma$ and $\rho_s$ (right). }\label{hor13}
\end{figure}

In this work, we will look into the derivation  of a static BH with a Dehnen-type DM halo exhibiting a quintessence field. We first construct the density profile of the Dehnen dark matter halo, which is a specific example of a double power-law profile defined by \cite{d1}: \begin{equation}
\rho =\rho _{s}\left( \frac{r}{r_{s}}\right) ^{-\sigma }\left[ \left( 
\frac{r}{r_{s}}\right) ^{\alpha }+1\right] ^{\frac{\sigma -\beta }{\alpha }},
\label{dens1}
\end{equation}%
where, $\rho _{s}$ and $r_{s}$ are  the central halo density radius.  Whereas  $\sigma$ determines the specific variant of the profile. The values of $\sigma$ lies within
$[0, 3]$ i.e. $\sigma = 3/2$ is used to fit the surface brightness profiles of elliptical galaxies which closely
resembles the de Vaucouleurs $r^{1/4}$ profile \cite{ds1}. In this paper, we use the parameters $\left( \alpha
,\beta ,\sigma \right) =\left( 1,4,0\right)$ \cite{dn14}. Therefore Eq. (\ref{dens1})
becomes 
\begin{equation}
\rho _{D}=\frac{\rho _{s}}{\left( \frac{r}{r_{s}}+1\right) ^{4}}.
\label{dens2}
\end{equation}
Our next step is to determine the mass distribution of the dark matter profile. Thus, the mass profile can be estimated using the relation as
\begin{equation}
M_{D}=4\pi \int\limits_{0}^{r}\rho \left( r^{\prime }\right) r^{\prime
2}dr^{\prime }=\frac{4\pi \rho _{s}r_{s}^{3}r^3}{3\left( r+r_{s}\right) ^{3}}.
\end{equation}
In a spherically symmetric spacetime, the mass distribution at the center of the dark matter halo can be used to calculate the tangential velocity of the particle moving within it:
\begin{equation}
v_{D}^{2}=\frac{M_{D}}{r}=\frac{4\pi \rho _{s}r_{s}^{3}r^2}{3r\left( r+r_{s}\right) ^{3}}.
\end{equation}
A pure dark matter halo can be described by a spherically symmetric line element:
\begin{equation}
ds^{2}=-F\left( r\right) dt^{2}+\frac{dr^{2}}{F\left( r\right) }+r^{2}\left(
d\theta ^{2}+\sin ^{2}\theta d\phi ^{2}\right).\label{m2}
\end{equation}%
where the metric function $F(r)$ is related to the tangential velocity through the
expression
\begin{equation}
v_{D}^{2}=r\frac{d}{dr}\left( \ln \sqrt{F(r)}\right) .
\end{equation}%
Solving for $F(r)$ we obtain 
\begin{equation}
F(r)=exp\left[- \frac{4\pi \rho _{s}r_{s}^{3}\left( 2r+r_{s}\right) }{%
3\left( r+r_{s}\right) ^{2}}\right] \approx 1-\frac{4\pi \rho
_{s}r_{s}^{3}\left( 2r+r_{s}\right) }{3\left( r+r_{s}\right) ^{2}},\label{mf1}
\end{equation}
where the leading order terms of the equation were retained. The final step is to combine the DM profile encoded in $F(r)$ with the BH metric function. To do so, we used Xu et al.'s formalism \cite{dn9}, which has also been used by others \cite{dn10,dn11,dn12,dn13,dn14}, to write the metric line element in this scenario. Therefore, the metric of a static and spherically symmetric BH embedded in DM halo in the background of  quintessential field is described by a line element \cite{Kiselev03,dn14} 
\begin{equation} 
ds^{2}=-f\left( r\right) dt^{2}+\frac{dr^{2}}{f\left( r\right) }+r^{2}\left(
d\theta ^{2}+\sin ^{2}\theta d\phi ^{2}\right) ,\label{m1}
\end{equation} where
\begin{equation}
f\left( r\right) =1-\frac{2M}{r}-\frac{4\pi \rho _{s}r_{s}^{3}\left(
2r+r_s\right) }{3\left( r+r_{s}\right) ^{2}}-\frac{\gamma}{r^{3\epsilon+1}},  \label{laps1}
\end{equation}
where $\gamma$ represents  the quintessential
state parameter and $\epsilon$
is the positive normalization factor depending on the density of the quintessence matter. 
We can explore the BH solution's behavior by taking advantage of curvature scalars. The  Kretschmann scalar for the metric (\ref{m1}) is
\[
R^{\mu \nu \alpha \beta }R_{\mu \nu \alpha \beta }=\frac{4\left(
r_{s}+r\right) ^{4}\left( 6M\left( r_{s}+r\right) ^{2}+3r^{-3\epsilon
}\left( r_{s}+r\right) ^{2}\gamma +4\pi r_{s}^{3}r\left( r_{s}+2r\right)
\rho _{s}\right) ^{2}}{9r^{6}\left( r_{s}+r\right) ^{8}}
\]%
\[
+\frac{\left( 12M\left( r_{s}+r\right) ^{3}+3r^{-3\epsilon }\left(
r_{s}+r\right) ^{4}\gamma \left( 1+3\epsilon \right) \left( 2+3\epsilon
\right) -32\pi r_{s}^{3} r^2(r_s+r)\rho + 24 r_{s}^{3} \pi r^3 (r_s +2r)\rho \right)^2}{9r^{6}\left( r_{s}+r\right) ^{8}}
\]%
\begin{equation}
+\frac{4\left( r_{s}+r\right) ^{2}\left( 6M\left( r_{s}+r\right)
^{3}+3r^{-3\epsilon }\left( r_{s}+r\right) ^{3}\gamma \left( 1+3\epsilon
\right) -8\pi r_{s}^{3} r^2(r_s +r)\rho +8 r_{s}^{3} \pi r^2 (r_s +2r)\rho \right)^2}{9r^{6}\left( r_{s}+r\right) ^{8}}.
\end{equation}
The Kretschmann scalar exhibits a singularity at $r = 0$. Thus, the BH solution's singularity at $r = 0$ is an essential singularity that no coordinate transformation can eliminate. \\ To visualize the metric function (\ref{laps1}), we display it as a function of $r$, as illustrated in figure \ref{lapse}. The graphic indicates that there are two horizons for the conjunction of the Dehnen type DM halo and the quintessence field. For large values of core density, there is no horizon that indicates a naked singularity.  In addition to event horizons, cosmological horizons become apparent when the quintessence term is taken into account. 
Spacetime horizons can be determined using  the condition $f(r) = 0$ i.e.,
\begin{equation}
 1-\frac{2M}{r}-\frac{4\pi \rho _{s}r_{s}^{3}\left(
2r+r_s\right) }{3\left( r+r_{s}\right) ^{2}}-\frac{\gamma}{r^{3\epsilon+1}}=0. \label{h9}  
\end{equation}
This equation (\ref{h9}) does not have an analytical solution. To derive the event and cosmological horizons, we shall use a numerical solution. As we proceed, we will focus on the special case $\epsilon=-2/3$.  The numerical results of these
two horizons are tabulated in table \ref{taba1}. 
\begin{center}
\begin{tabular}{|c|c|c|c|c|c|c|}
 \hline 
 & \multicolumn{3}{|c|}{ $\rho_s =0.8$}&\multicolumn{3}{|c|}{ $r_{s}=0.4$}
\\ \hline 
$\gamma $ & $r_{s}$ & $r_{h}$ & $r_{c}$ & $\rho
_{s}$ & $r_{h}$ & $r_{c}$ \\ \hline
$0.01$ & $0.2$ & $2.0905$ & $97.9026$ & $0.2$ & $2.12843$ & $97.8471$ \\ 
& $0.6$ & $3.22949$ & $96.4389$ & $0.6$ & $2.30714$ & $97.6238$ \\ 
& $1$ & $8.36711$ & $90.506$ & $1$ & $2.49217$ & $97.3995$ \\ 
$0.02$ & $0.2$ & $2.13839$ & $47.8547$ & $0.2$ & $2.17849$ & $47.797$ \\ 
& $0.6$ & $3.36132$ & $46.3073$ & $0.6$ & $2.3674$ & $47.5636$ \\ 
& $1$ & $9.6502$ & $39.2229$ & $1$ & $2.56399$ & $47.3277$ \\ 
$0.03$ & $0.2$ & $2.19111$ & $31.1353$ & $0.2$ & $2.23373$ & $31.0752$ \\ 
& $0.6$ & $3.51788$ & $29.4842$ & $0.6$ & $2.43463$ & $30.8298$ \\ 
& $1$ & $13.4464$ & $18.7599$ & $1$ & $2.64503$ & $30.5801$%
\\ 
 \hline
\end{tabular}
\captionof{table}{Numerical results for the event horizon $r_h$ and cosmological horizon $r_c$ of the Dehnen-type DM BH in the background of quintessence field.} \label{taba1}
\end{center}
Table \ref{taba1} shows that a Dehnen-type DM halo with two parameters (core radius $r_s$ and core density $\rho_s$) has similar effects on the event horizon, increasing $r_h$ and decreasing $r_c$. (see figure \ref{hor12}). However, the event horizon is more influenced by core radius than density.  The cosmological horizon reduces dramatically with increasing the quintessence field value, but only marginally with $\rho_s$ and $r_s$  (see figure \ref{hor13}).

\section{Shadow} \label{sec3}

\begin{figure}
  \begin{center}
\includegraphics[scale=0.55]{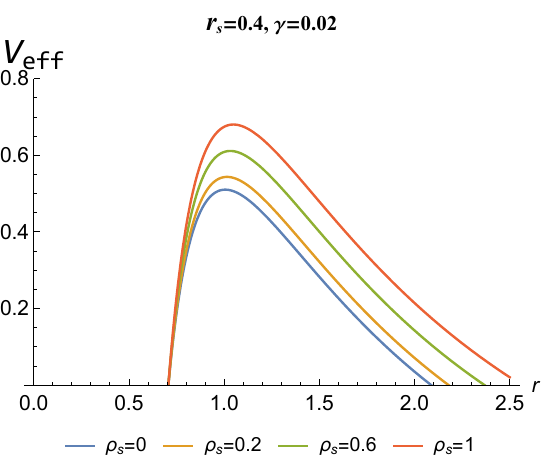}
\includegraphics[scale=0.55]{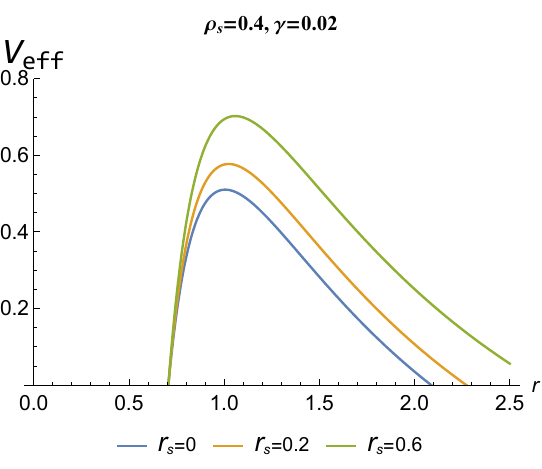} \includegraphics[scale=0.55]{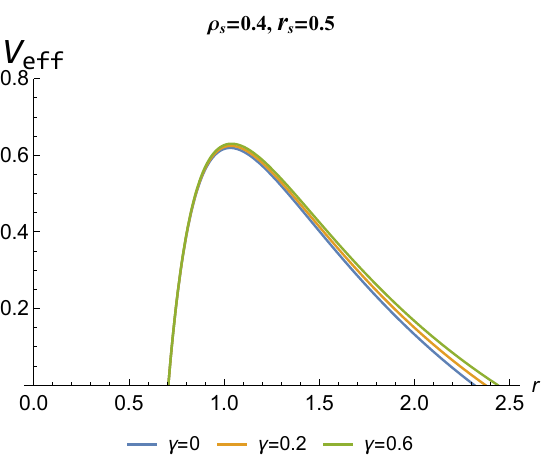}
\end{center}
\caption{Behaviours of effective potential $V_S$  with respect to radial distance $r$ for different values of core density, core radius and the quintessence parameter. }\label{new1}
\end{figure}

\begin{figure}
    \centering
    \includegraphics[width=0.45\linewidth]{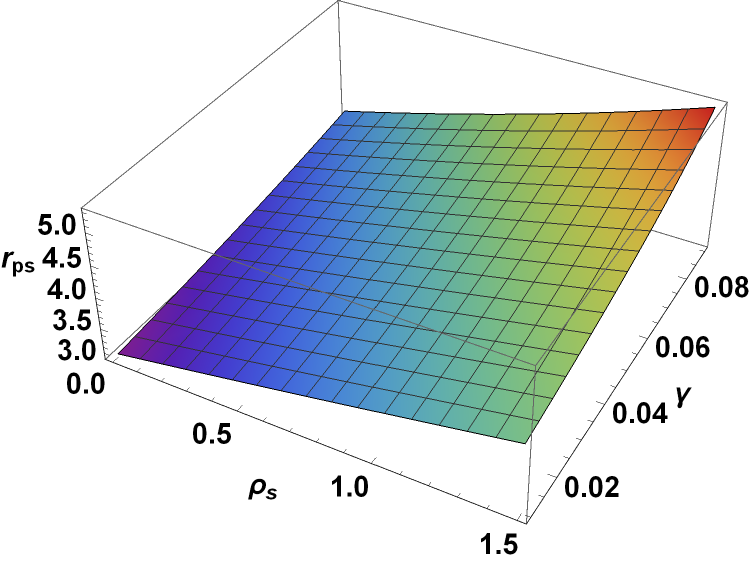} \includegraphics[width=0.45\linewidth]{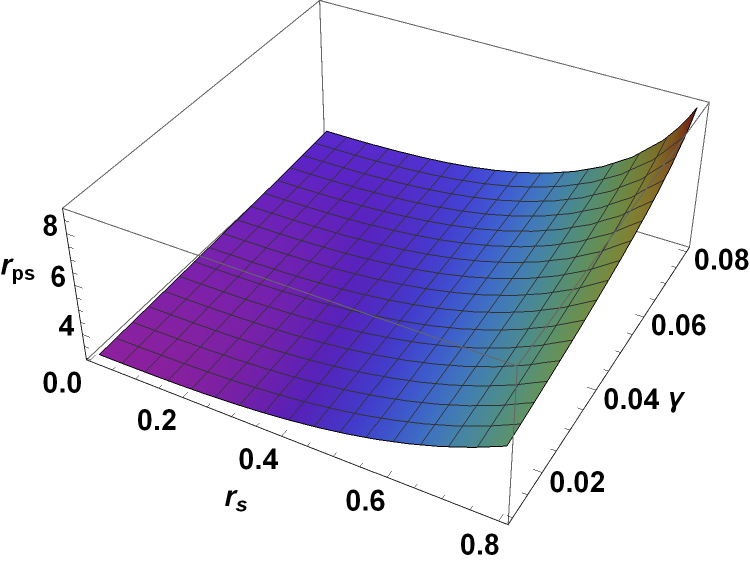}
    \caption{Variation of photon sphere radius with $\rho_s$ and $\gamma$ keeping $r_s = 0.4$ (left) and with $r_s$ and $\gamma$ keeping $\rho_s = 0.5$ (right)}
    \label{ps22}
\end{figure}

\begin{figure}
\begin{center}
\includegraphics[scale=0.55]{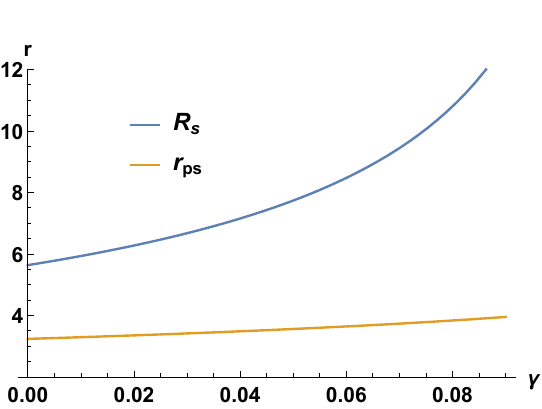}\includegraphics[scale=0.55]{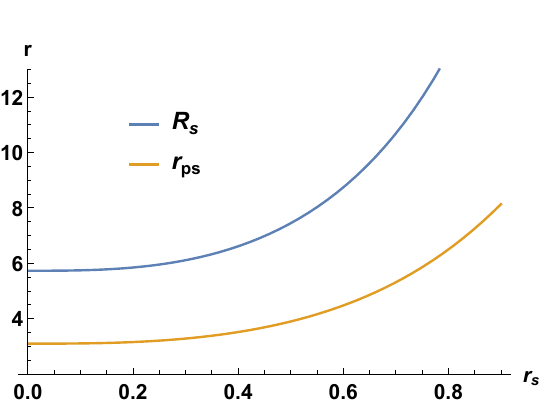}\includegraphics[scale=0.55]{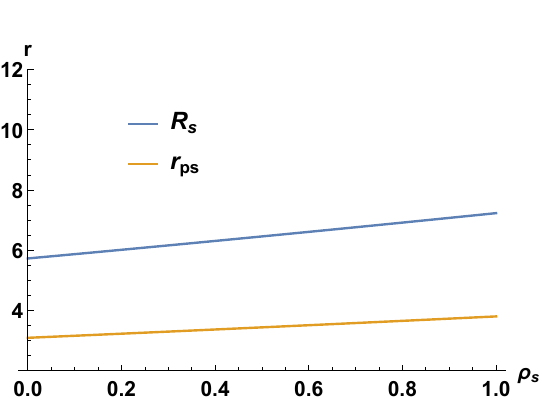}
\end{center}
\caption{Variation of $r_{ps}$ and $R_s$ with quintessence parameter keeping $r_s = 0.5$ and $\rho_s=0.2$ (left), with core radius keeping $\gamma = 0.02$ and $\rho_s=0.6$ (middle) and with core density keeping $\gamma = 0.02$ and $r_s=0.4$(right) .}\label{ps24}
\end{figure}

\begin{figure}
\begin{center}
\includegraphics[scale=0.4]{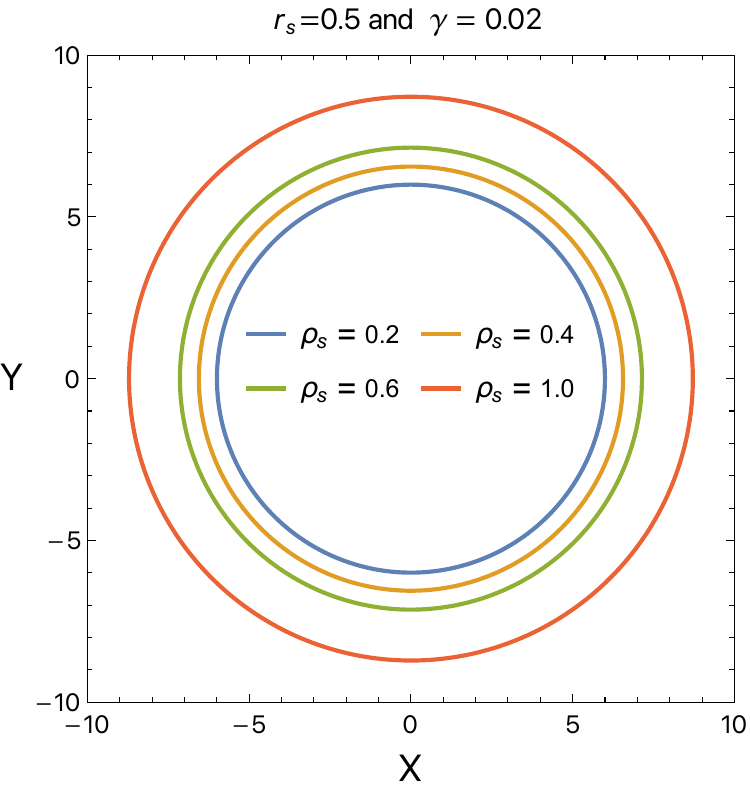}\includegraphics[scale=0.4]{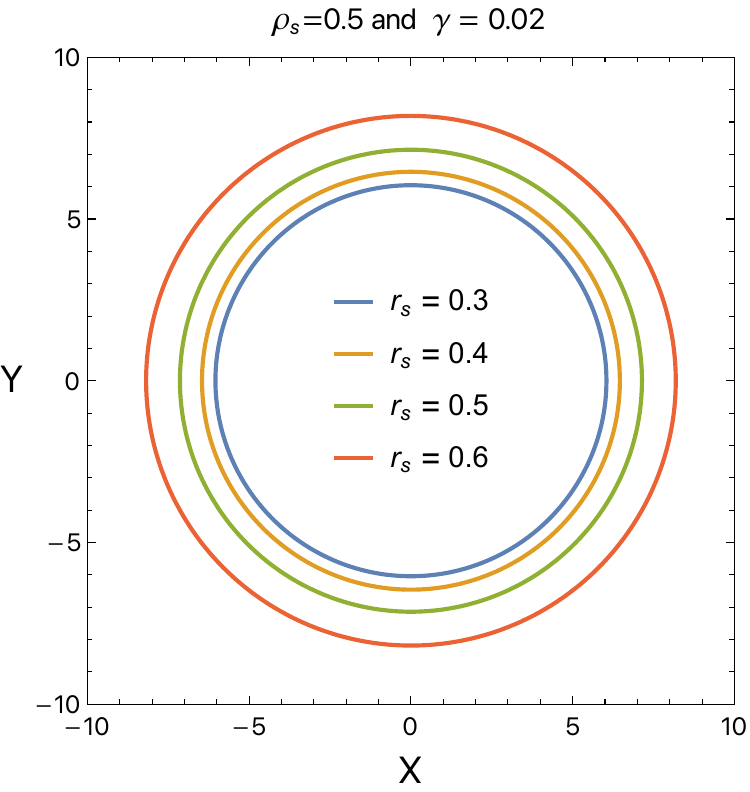}\includegraphics[scale=0.4]{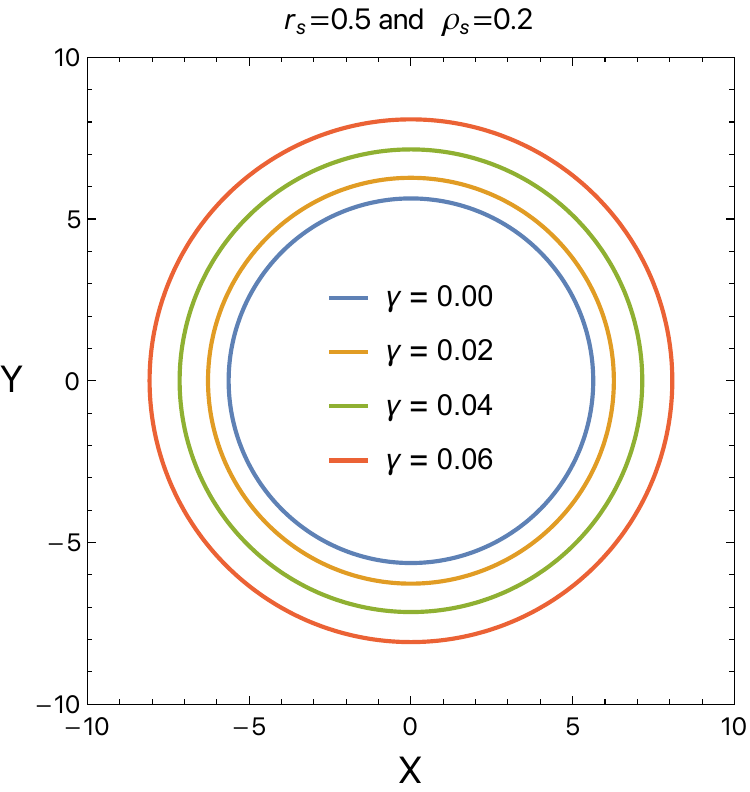}
\end{center}
\caption{The BH shadow profile for different values of the DM density $\rho_s$ (left) and the core radius $r_s$ (middle) and the quintessence parameter $\gamma$ (right).} \label{sh_3cases}
\end{figure}

Shadow analysis relies on null geodesics, which is the motion of photons around black holes. Thus, the Lagrangian in the
background of the Dehnen-type DM BH surrounding with quintessence field is given by 
\begin{equation}
\mathcal{L}(x,\dot{x})=\frac{1}{2}\,g_{\mu\nu}\dot{x}^{\mu}\dot{x}^{\nu}.
\end{equation} 
 where over-dot is differentiation with respect to affine parameter $\tau$ and we confine ourselves to the
equatorial plane $(\theta=\pi/2)$.  With the metric independent of both $t$ and $\phi$, we get two conserved quantities of motion – energy $E$ and angular momentum $L$. 
\begin{equation}
 E=f(r)\dot{t}, \hspace{1cm} L=r^2 \dot{\phi}   
\end{equation} 
 Thus, we obtain the following equation for radial coordinates:
 \begin{equation}
     \dot{r}^2=E^2
-f(r)\frac{L^2}{r^2}=E^2-V_{eff}(r)
\end{equation}
 where the effective potential 
 \begin{equation}
    V_{eff}=\frac{f(r)}{r^2}\left( \frac{L^2}{E^2}-1\right). \label{effpo1}
\end{equation} 
Figure \ref{new1} shows the behavioure of the effective potential with respect to radial distance $r$. It demonstrates that  potential peak values increase with increasing  DM halo parameters ($\rho_s, r_s)$ and quintessence parameter $\gamma$.
\\  The radius of an unstable photon orbit can be obtained using the following conditions $V_{eff}=0=V'_{eff}$. To determine the BH shadow radius, we only need the photon sphere radius and critical impact parameter. The equation of photon sphere radius is given by \begin{equation}
    r_{ps}f'(r_{ps})-2f(r_{ps})=0. \label{ps8}
\end{equation}
A critical impact parameter $b_{ps}$ corresponding to the photon orbit, also known as the shadow radius $(R_s)$ of the BH, can be identified as follows:
\begin{equation}
    b_{ps}=R_s=\frac{r_{ps}}{\sqrt{f(r_{ps})}}.
\end{equation}
Substituting the metric function (\ref{laps1}) in Eq. (\ref{ps8}) we obtain 
\begin{equation}
    (18M+r(3r\gamma -6))(r_s +r)^3-8r^3_{s}\pi(r^2_{s}+3r_{s}r+3r^2)\rho_s=0. \label{ps2}
\end{equation}
We observe that photon radius cannot be expressed analytically. We will thus obtain its values numerically. Table \ref{taba2} shows numerical values for the photon sphere, and consequently the shadow radius. In table \ref{taba2}, we examined the impact of the quintessence parameter $\gamma$ and the Dehnen-type dark matter halo with two parameters $r_s$ and $\rho_s$ on the BH  shadow.
\begin{center}
\begin{tabular}{|c|c|c|c|c|c|c|}
 \hline 
 & \multicolumn{3}{|c|}{ $\rho_s =1$}&\multicolumn{3}{|c|}{ $r_{s}=0.4$}
\\ \hline 
$\gamma $ & $r_{s}$ & $r_{ps}$ & $R_{s}$ & $\rho
_{s}$ & $r_{ps}$ & $R_{s}$ \\ \hline
$0.01$ & $0.2$ & $3.1382$ & $5.61977$ & $0.2$ & $3.17799$ & $5.70235$ \\ 
& $0.4$ & $3.72768$ & $6.77653$ & $0.6$ & $3.44847$ & $6.23092$ \\ 
& $0.6$ & $5.34526$ & $10.0413$ & $1$ & $3.72768$ & $6.77653$ \\ \hline
$0.02$ & $0.2$ & $3.19095$ & $5.92313$ & $0.2$ & $3.23244$ & $6.0157$ \\ 
& $0.4$ & $3.80555$ & $7.23178$ & $0.6$ & $3.51394$ & $6.6105$ \\ 
& $0.6$ & $5.52092$ & $11.1272$ & $1$ & $3.80555$ & $7.23178$ \\ \hline
$0.03$ & $0.2$ & $3.24751$ & $6.27496$ & $0.2$ & $3.29088$ & $6.38009$ \\ 
& $0.4$ & $3.89042$ & $7.77988$ & $0.6$ & $3.58475$ & $7.05941$ \\ 
& $0.6$ & $5.72208$ & $12.6021$ & $1$ & $3.89042$ & $7.77988$%
\\ 
 \hline
\end{tabular}
\captionof{table}{Numerical results for the photon sphere $r_{ps}$ and shadow radius $R_s$ of the Dehnen-type DM BH in the background of quintessence field.} \label{taba2}
\end{center}
 To visualize the effects of the BH parameters we present the figures \ref{ps22} and \ref{ps24}. It demonstrates that both the photon sphere and the shadow radius increase with core density and core radius. Both radii increase roughly linearly with core density, but not linearly with core radius. Similarly, as the quintessence parameter increases, so do both radii. We infer that the presence of Dehnen type DM and quintessence field enhances the size of the shadow. This behavior is also reflected in Fig. \ref{sh_3cases}. As can be seen from Fig.~\ref{sh_3cases} the shadow size eventually grows with increasing DM parameters and quintessence parameter $\gamma$. 

We then turn to provide insights into the BH parameters using recent EHT observations \cite{Akiyama19L1,Akiyama19L6}). For that we explore the obtained theoretical results to obtain precise constraints on these parameters.  For the astrophysical implications, 
two sources, Sgr A$^{\star}$ and M87$^{\star}$ which can provide upper constraints for the BH parameters considered here, are assumed to be candidates to describe static and spherically symmetric BHs for our theoretical approach, supporting all assumptions. Taken EHT observational data, we are able then to obtain upper limits of the DM halo density $\rho_s$ and the quintessence parameter $\gamma$.  For our approach, we adapt the observational shadow's angular diameter $\theta$, the distance $D$, and the mass of Bhs at the center of Sgr A$^{\star}$ and M87$^{\star}$ galaxies. The recent EHT observations have reported for M87$^{\star}$ and Sgr A$^{\star}$ as follows: \begin{eqnarray}
\theta_{M87^{\star}}=42 \pm 3 \mu  
\, \, \mbox{and}\, \,  \theta_{Sgr A^{\star}}=48.7 \pm 7 \mu\, ,
\end{eqnarray}
and distances 
$D_{M87^{\star}}= 16.8 \pm 0.8 M pc$ and $D_{M87^{\star}}=8277 \pm 9 \pm 33 pc$ between Earth and M87$^{\star}$ and Sgr A$^{\star}$ with masses $M_{M87^{\star}} = (6.5 \pm 0.7) \times 10^9 M_{\odot}$ and  $D=8277 \pm 9 \pm 33 pc$ and $M_{Sgr A^{\star}} = (4.297 \pm 0.013) \times 10^6 M_{\odot}$ (see, e.g., in Refs. \cite{Akiyama19L1,Akiyama19L6}). One can then evaluate the shadow diameter per unit mass by adapting the equation given by 
\begin{eqnarray}
    d_{sh}=\frac{D\,\theta}{\,M}\, ,
\end{eqnarray}
\begin{figure}
    \centering
    \includegraphics[scale=0.55]{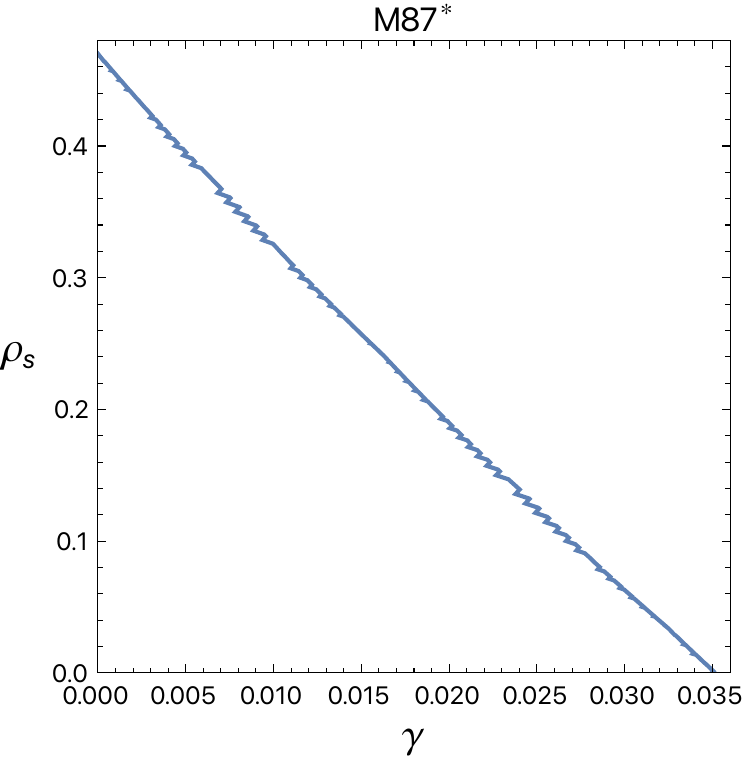}%
    \includegraphics[scale=0.55]{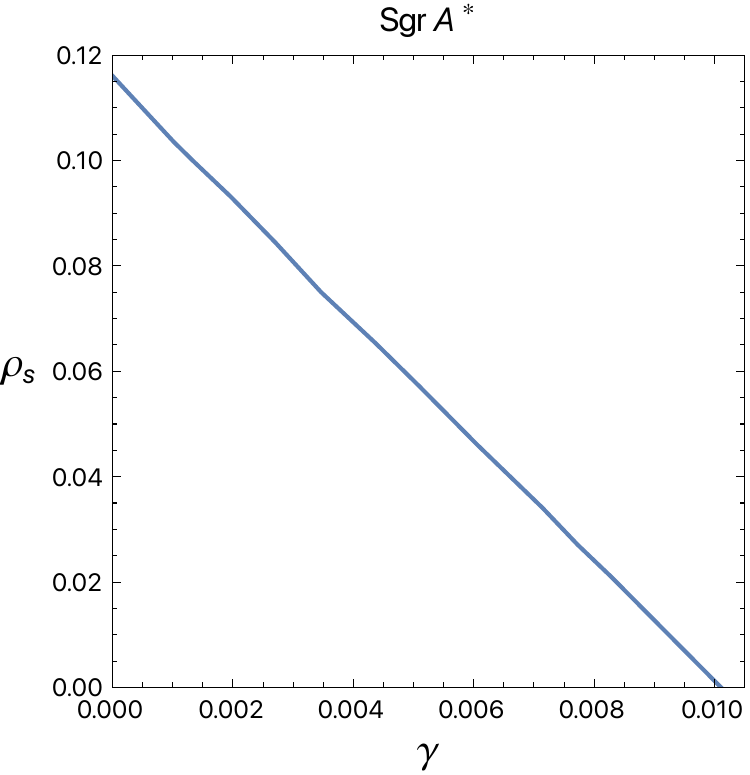}
    \caption{The upper values of the DM density $\rho_s$ and quintessence parameter $\gamma$ for M87$^{\star}$ and Sgr A$^{\star}$. Here, we have set $r_s=0.5$ and $M=1$. }
    \label{fig:constraint}
\end{figure}
which corresponds to $d^{M87^{\star}}_{sh}=(11 \pm 1.5)M$ for M87$^{\star}$ and $d^{Sgr^{\star}}_{sh}=(9.5 \pm 1.4)M$ for Sgr A$^{\star}$, respectively. Taking these values into consideration for Sgr A$^{\star}$ and M87$^{\star}$, we obtain the possible values of $\rho_s$ and $\gamma$, which  are demonstrated in Fig.~\ref{fig:constraint}. From observational data of M87$^{\star}$, one can deduce that $\rho_s$ and $\gamma$ may have larger values compared to those from Sgr A$^{\star}$. Subsequently, EHT observational data for M87$^{\star}$ and Sgr A$^{\star}$ can provide precise constraints on the DM density parameter $\rho_s<0.48,\, 0.12$ and the quintessence parameter $\gamma<0.035,\, 0.01$, respectively, for the Dehnen-type DM BH system with the quintessence
field background. 

\section{deflection angle} \label{sec4}

Unlike previous section, here, we now determine the deflection angle around Schwarzschild BHs embedded in a Dehnen type DM halo with quintessential background field by adapting the method as a new thought experiment developed by Gibbons and Werner. This method refers to the Gauss-Bonnet theorem (GBT), evaluating the deflection angle for spherically symmetric spacetimes, (see, e.g., \cite{Gibbons08CQG,Werner12GBT}). It is to be emphasized that the GBT method was extended to the axisymmetric spacetimes \cite{Ono17GBT}, non-asymptotically flat spacetimes \cite{Ishihara16GBT,Ishihara16}. 
To evaluate the weak deflection angle, this thought experiment has since been widely extended to various various situations and gravity models \cite{Jusufi18GBT,Li20,Zhang21GBT,DCarvalho21,isMandal:2023eae,Al-Badawi24EPJC_GBTa,Al-Badawi24EPJC_GBTb}. It is worth noting, however, that we consider the weak deflection angle in the background field of the BH embedded in a Dehnen type DM halo surrounded by the quintessential field. Taking the GBT method into consideration, we need to determine the weak deflection angle by restricting null geodesics for photon motion. For that we use the optical metric for BH spacetime, which is written as follows:  
\begin{eqnarray}\label{opt:metric}
\mathrm{d}\sigma^{2} = g_{kl}^{\mathrm {opt}}\mathrm{d}x^{k} \mathrm{d}x ^{l}
= dr_{*}^2+\mathcal{F}^2(r_{*})d\phi^2\, ,
\end{eqnarray}
where 
\begin{equation} \label{Eq:tor1}
\mathcal{F}\big(r_{*}(r)\big)=\frac{r}{\sqrt{f(r)}},
\end{equation}
with $f(r)=\left(1-{2M}/{r}-{4\pi \rho _{s}r_{s}^{3}\left(
2r+r_s\right) }/{3\left( r+r_{s}\right) ^{2}}-{\gamma}/{r^{3\epsilon+1}} \right)$
and $r_{*}$ which refers to the tortoise coordinate and is given by 
\begin{equation}\label{Eq:tor2}
r_{*}=\int{\frac{\mathrm{d}r}{1-\frac{2M}{r}-\frac{4\pi \rho _{s}r_{s}^{3}\left(
2r+r_s\right) }{3\left( r+r_{s}\right) ^{2}}-\frac{\gamma}{r^{3\epsilon+1}}}}\, .
\end{equation} 
Within the optical metric Eq.~(\ref{opt:metric}), we further evaluate the Gaussian curvature $\cal{K}$ by determining the non-vanishing components of Christoffel symbols associated with the BH spacetime \cite{Wald:1984}. We then write non-vanishing components of Christoffel symbols as  
\begin{eqnarray} \label{Eq:Chris1}
\Gamma_{\phi \phi}^{r_{*}}&=&-\mathcal{F}(r_{*})\frac{\mathrm{d}\mathcal{F}(r_{\star})}{\mathrm{d}{r_{\star}}},\\
\Gamma_{r_{*}\phi}^{\phi}&=&\frac{1}{\mathcal{F}(r_{*})}\frac{\mathrm{d}\mathcal{F}(r_{\star})}{\mathrm{d}{r_{\star}}}, \label{Eq:Chris2}
\end{eqnarray}
with the determinant $\det\tilde{g}_{kl}=\mathcal{F}^{2}(r^{\star})$. Taking all together the Gaussian curvature $\mathcal{K}$ can be defined by \cite{Mandal:2023} 
\begin{eqnarray}
\mathcal{K}=-\frac{R_{r_{*}\phi r_{*}\phi}}{\det\tilde{g}_{r\phi}}=-\frac{1}{\mathcal{F}(r_{\star})}\frac{\mathrm{d}^{2}\mathcal{F}(r_{\star})}{\mathrm{d}{r_{\star}}^{2}}\, .
\end{eqnarray}

It must also be noted that the Gaussian curvature $\mathcal{K}$ can be found alternatively within the context of the variable $r$ \cite{Gibbons08CQG,Chandrasekhar:1985,Sakalli:2016}, i.e., it consequently reads as 
\begin{eqnarray}\label{Eq:GK}
\mathcal{K} & = &-\frac{1}{\mathcal{F}(r^{\star})}\left[\frac{\mathrm{d}r}{\mathrm{d}r^{\star}}\frac{\mathrm{d}}{\mathrm{d}r}\left(\frac{\mathrm{d}r}{\mathrm{d}r^{\star}}\right)\frac{\mathrm{d}f(r)}{\mathrm{d}r}+\left(\frac{\mathrm{d}r}{\mathrm{d}r^{\star}}\right)^{2}\frac{\mathrm{d}^{2}f(r)}{\mathrm{d}r^{2}}\right]\, .
\end{eqnarray}
Here, recalling Eq.~\eqref{Eq:GK} and employing Eqs.~\eqref{Eq:tor1} and \eqref{Eq:tor2} we derive the Gaussian curvature with the equatorial plane area $\mathrm{d}S=\sqrt{|\det{g}_{kl}^{opt}|}\mathrm{d}r \mathrm{d}\phi$ in the optical metric as follows:  
\begin{eqnarray}
\mathcal{K}\mathrm{d}S&=&
-\frac{9 \big[9 M^2 (r_s + r)^4 + 12 r_s^3 \pi r^3 \big(2 M r - r_s (r-4 M)\big) \rho_s + 
   16 r_s^6 \pi^2 r^4 \rho_s^2\big]\lambda}{2 \big(3 (2 M - r) (r_s + r)^2 + 
   4 r_s^3 \pi r (r_s + 2 r) \rho_s\big)^2 \big(
 9 - \frac{18M}{r} - \frac{12 r_s^3 \pi (r_s + 2 r) \rho_s}{(r_s + r)^2}\big)^{1/2}} \nonumber\\&+&
\Big(48 \pi  r_s^3 r \rho_s (r_s+r)^2 \left(2 r_s^3 M+8 r_s^2 M r+r^3 (r_s+6 M)+6 r_s M r^2-2 r^4\right)\nonumber\\&&-64 \pi ^2 r_s^6 r^4 \rho_s^2 \left(r_s^2-3 r^2\right)+36 M (r_s+r)^6 (3 M-2 r) \Big)\nonumber\\&\times&
\frac{4 \big(9-\frac{18 M}{r}-\frac{12 \pi  r_s^3 \rho_s (r_s+2 r)}{(r_s+r)^2}\big)^{-1/2}}{r^2 (r_s+r)^4 \Big(3 (r_s+r)^2 (r-2 M)-4 \pi  r_s^3 r \rho (r_s+2 r)\Big)}\,\mathrm{d}r\,\mathrm{d}\phi\, .
\end{eqnarray}
Taking the contribution of the Dehnen-type DM BH spacetime to the Gaussian curvature, we write the geodesic curvature as  \cite{Gibbons08CQG,Al-Badawi24EPJC_GBTa}
\begin{eqnarray}
\frac{\mathrm{d}\sigma}{\mathrm{d}\phi}\bigg|_{C_{R}}=
\left( \frac {r^2}{f(R)} \right)^{1/2}\, .
\end{eqnarray}
Here, we note that $C_{R}$ refers to a curve defined by $r(\phi) = R = constant$ with its radius $R$. To determine the deflection angle using the GBT we further consider the limiting case ${R} \rightarrow \infty$ that results in having the relation $\theta_{0}+\theta_{S}=\pi$ between the object and source.  As a consequence, the above equation in the limiting case can be rewritten as follows:  
\begin{eqnarray}
\lim_{R\to\infty} \kappa_g\frac{\mathrm{d}\sigma}{\mathrm{d}\phi}\bigg|_{C_R}\approx 1\, .
\end{eqnarray}
Taking spatial infinity into consideration $R\to\infty$ and imposing the straight
light approximation $r=b/\sin\phi$, the equation for determining the small deflection angle $\tilde{\delta}$ using the GBT method reads as follows \cite{Gibbons08CQG,Al-Badawi24EPJC_GBTa}:
\begin{eqnarray}
 \int^{\pi+\tilde{\delta}}_0 \left[\kappa_g\frac{\mathrm{d}\sigma}{\mathrm{d}\phi}\right]\bigg|_{C_R}\mathrm{d}\phi-\pi 
 =-\lim_{R\to\infty}\int^\pi_0\int^{\infty}_{\frac{b}{\sin\phi}}\mathcal{K}\,\mathrm{d}S\, .
\end{eqnarray}
It should be noted that we have defined $b$ as the impact parameter in the equation of the deflection angle. Consequently, the deflection angle around the Dehnen-type DM BH with a background of quintessence field in the weak
limit approximation can be computed explicitly by the following approximate form
\begin{eqnarray}\label{Eq:def_angle}
\tilde{\delta} \approx
\frac{4 M}{b}-\frac{4 \pi  \gamma r_s^4 \rho_s}{b}+\frac{8 \pi  \gamma r_s^3 M \rho_s}{b}+\frac{3 \gamma M^2}{b}+\frac{\pi  \gamma M}{2}+\frac{16 \pi  r_s^3 \rho_s}{3 b}+\frac{16 \pi ^2 \gamma r_s^6 \rho_s^2}{3 b}\, .
\end{eqnarray}
\begin{figure*}
    \centering
{\includegraphics[width=7.75cm]{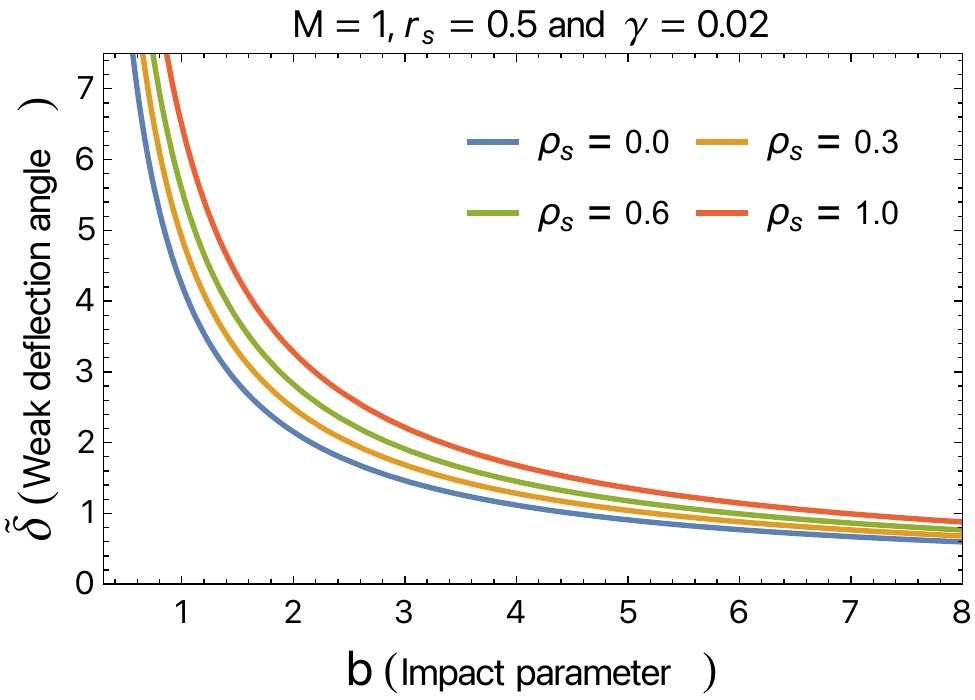} }\qquad
    {{\includegraphics[width=7.75cm]{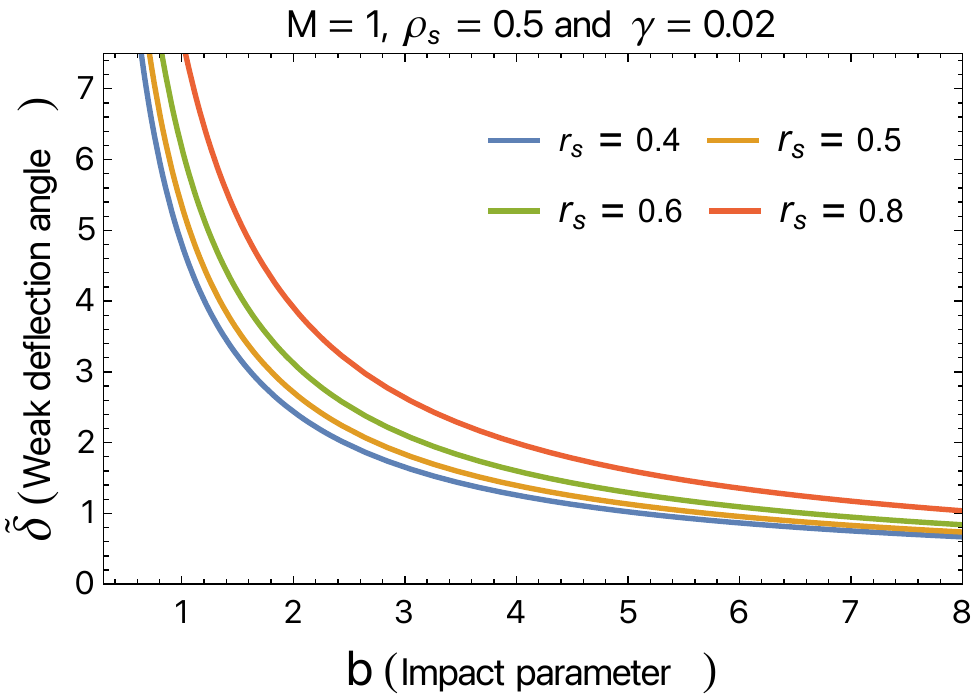}}}\qquad

    {{\includegraphics[width=7.75cm]{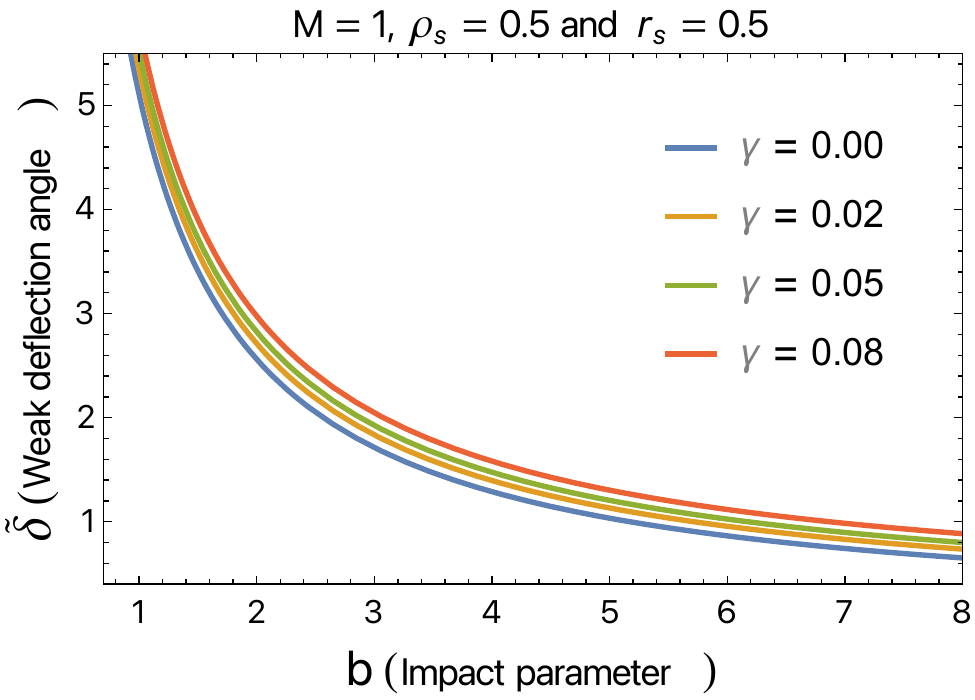}}}
    \caption{The profile of the deflection angle around the Dehnen-type DM BH with a background of quintessence field  as a function of impact parameter $b$ for different values of $\rho$ (top left panel), $rs$ (top right panel) and $\gamma$ (bottom panel). }
    \label{fig:def_an}
\end{figure*}
As can be seen from Eq.~(\ref{Eq:def_angle}), the deflection angle is influenced by the quintessential field parameter $\gamma$ and the DM halo density $\rho_s$ and the halo core radius $r_s$. We then analyze the deflection angle within the context of the optical metric for the BH spacetime considered here, resulting in a deeper understanding of the impact of BH parameters on the deflection angle in the weak form. In Fig~\ref{fig:def_an}, we depict the deflection angle profile against the impact parameter $b$. As seen in Fig~\ref{fig:def_an}, the left and right panels in the top row reflect the role of the DM density parameter $\rho_s$ and the halo core radius $r_s$ on the deflection angle profile while keeping the quintessence field parameter $\gamma$ fixed. The bottom panel reflects the role of the parameter $\gamma$ when keeping $\rho_s$ and $r_s$ constant for various cases. It is clearly seen from Fig.~\ref{fig:def_an} that the deflection angle is inversely proportional to the impact parameter $b$. Interestingly, we observe that the deflection angle has a similar changing rate, resulting in an increase due to the impact of the parameters $\rho_s$ and $r_s$. Similarly, the role of the quintessence parameter $\gamma$ has the same influence on the deflection angle compared to the effects of $\rho$ and $r_s$. It is important to note that the impact of the DM parameters $\rho_s$ and $r_s$ on the deflection angle is much more prominent around the BH compared to a distance away from the BH. This happens because the Dehnen-type DM profile is distributed well around the BH. Taken altogether, one can infer that these parameters, $\rho_s$ and $r_s$ together with the quintessence parameter $\gamma$, have similar effects that can be interpreted as repulsive gravitational charges,  causing the deflection angle to increase to possible larger values. For being for more informative, we next intend to consider quasinormal modes to explore the unique aspects of the Dehnen-type DM BH spacetime with a background of the quintessence field. 

\section{quasinormal modes} \label{sec5}

\begin{figure}[H]
\begin{center}
\includegraphics[scale=0.55]{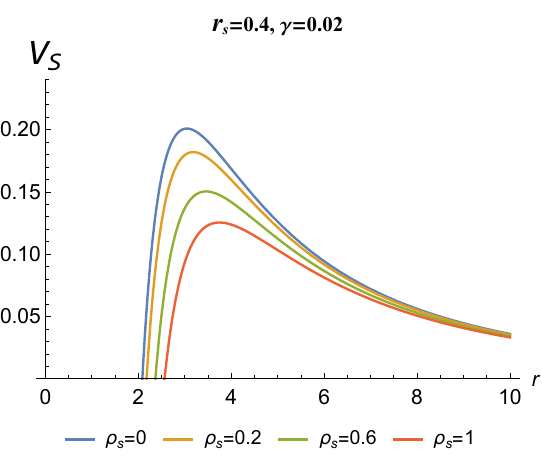}
\includegraphics[scale=0.55]{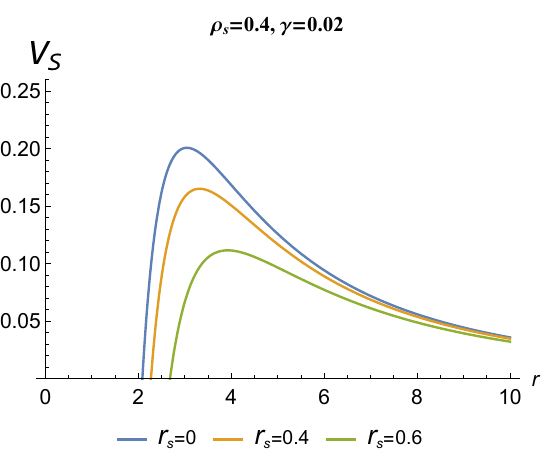} \includegraphics[scale=0.55]{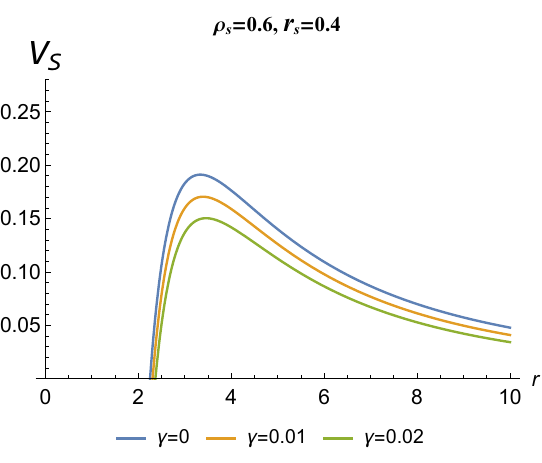}
\end{center}
\caption{Behaviours of scalar potential $V_S$  with respect to radial distance $r$ for different values of core density, core radius and the quintessence parameter. }\label{fpotS}
\end{figure}

\begin{figure}[H]
\begin{center}
\includegraphics[scale=0.55]{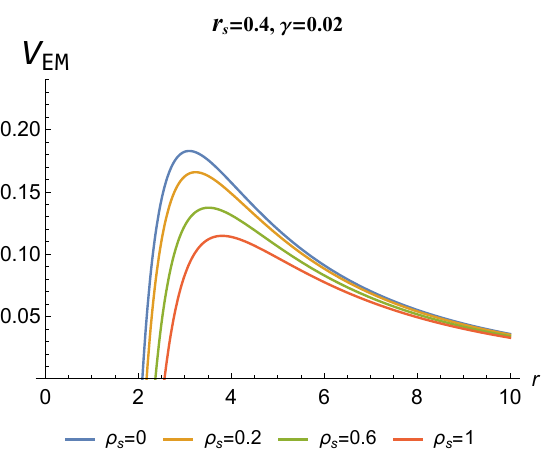}
\includegraphics[scale=0.55]{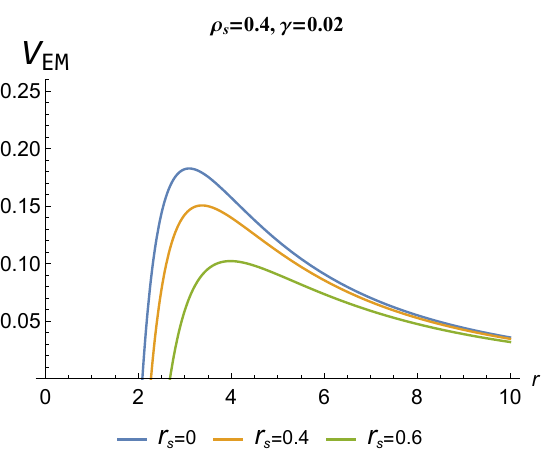} \includegraphics[scale=0.55]{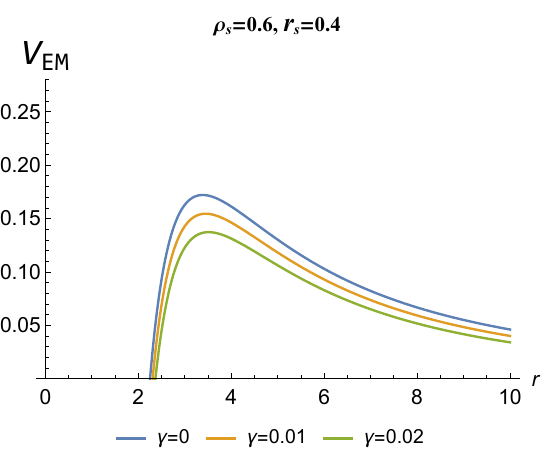}
\end{center}
\caption{Behaviours of scalar potential $V_{EM}$  with respect to radial distance $r$ for different values of core density, core radius and the quintessence parameter. }\label{fpotE}
\end{figure}

In this section, we will consider the scalar and electromagnetic 
fields perturbations of the Dehnen-type DM BH in the background of quintessence field to investigate the behaviour of QNMs. The gravitational waves emitted by BH coalescence provide valuable information about the nature of spacetime and are independent of any given initial stage. They also provide information about the BH system's stability in the presence of perturbations. The QNMs of scalar and electromagnetic field perturbations are wave equation solutions that fulfil particular boundary requirements both near the BH horizon and away from it. The solution must meet the requirements for purely ingoing waves at the event horizon and purely outgoing waves at the cosmological horizon or spatial infinity. \\  Since the existing system has spherical symmetry, then the general master equation for computing QNMs is:
\begin{equation}
\frac{\text{d}^{2}\Psi}{\text{d}r_{\ast }^{2}}+\left( \omega^{2}-V(r_{\ast })\right) \Psi=0,  \label{s3}
\end{equation}
where the tortoise coordinate $r_{\ast }$ is defined by $r_{\ast } = \int \frac{dr}{f}$. The effective potentials $V(r_{\ast })$ for scalar ($V_S$) and electromagnetic ($V_{EM}$) types of perturbation fields in Eq. (\ref{s3}) are  
\begin{equation}
V_{S}(r)=f \left( \left(\frac{1}{2}+l\right)\frac{1 }{r^{2}}+\frac{f^{\prime }}{r} \right), \label{pots}
\end{equation}
\begin{equation}
V_{EM}(r)=\left(\frac{1}{2}+l\right)\frac{f }{r^{2}},  \label{pots10} 
\end{equation}
where $l$ is the multipole moment. 
Figures \ref{fpotS} and \ref{fpotE} show graphs of the potentials (\ref{pots}) and (\ref{pots10}) to explore both the effect of the DM halo density as well as the quintessence parameter. The figures show that the DM halo (core density and radius parameters) and the quintessence parameter both have a large impact on the potentials, albeit in similar ways.  Potential peak values decrease with increasing  DM halo parameters ($\rho_s, r_s)$ and quintessence parameter $\gamma$. In comparison with the scalar (\ref{pots10}) and EM (\ref{pots}) potentials, the BH parameters have opposite effects on the effective potential (\ref{effpo1}). It is worth noting that, the effects of the neglected terms in the metric function (\ref{mf1}) on the peak of the perturbation potential are small enough to lead to variations in the QNMs spectrum.  

\begin{figure}
    \centering
    \includegraphics[width=18cm]{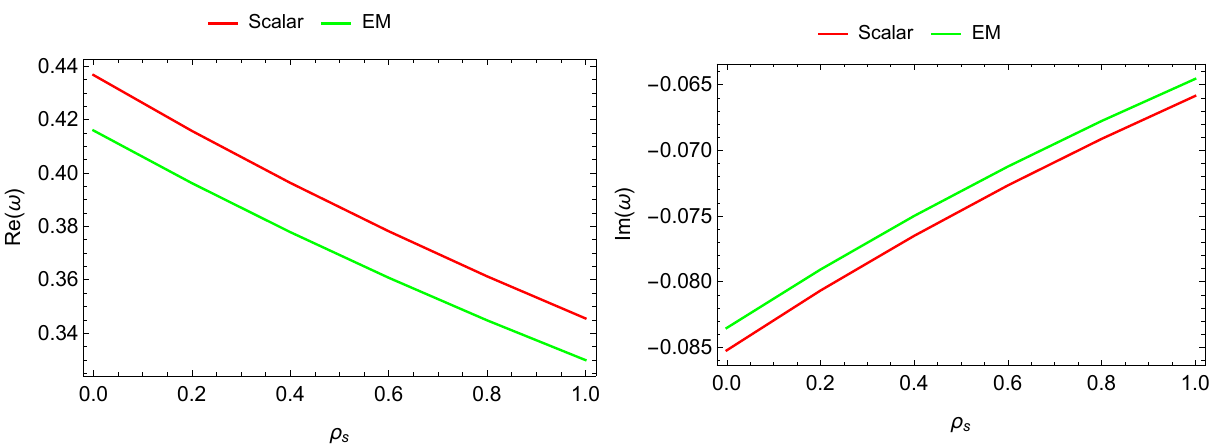}
    \caption{ Variation of amplitude and damping of QNMs with respect to the central density of the DM halo  parameter  for scalar and EM perturbations.}
    \label{realRo}
\end{figure}

\begin{figure}
    \centering
    \includegraphics[width=18cm]{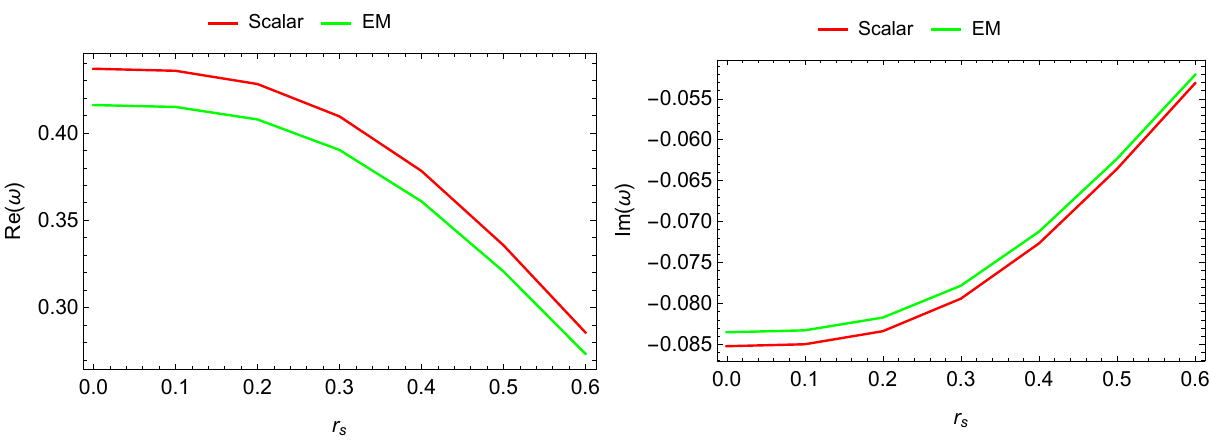}
    \caption{ Variation of amplitude and damping of QNMs with respect to the central radius of the DM halo  parameter  for scalar and EM perturbations.}
    \label{realS}
\end{figure}

\begin{figure}
    \centering
    \includegraphics[width=18cm]{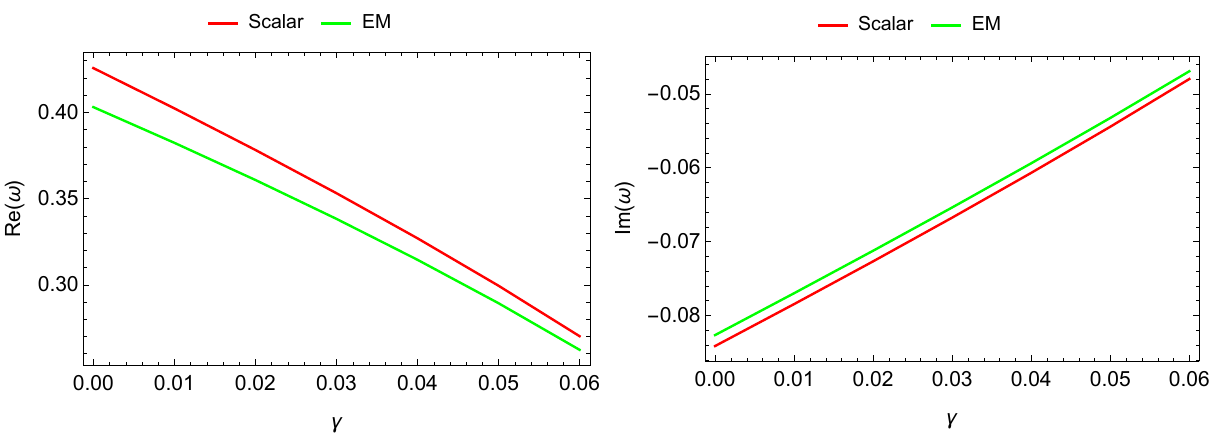}
    \caption{ Variation of amplitude and damping of QNMs with respect to the quintessence  parameter  for scalar and EM perturbations.}
    \label{realG}
\end{figure}

The WKB approximation approach is commonly used for calculating QNMs. It was initially introduced by Iyer \cite{d3}¸ and later expanded to higher levels by Konoplya \cite{d4}. The WKB approach is successful for low overtone numbers $n$, particularly for $n <l$.\\ Using the effective potentials, we determine numerically the quasinormal frequencies for scalar and electromagnetic perturbations using the 6th order WKB approximation. Tables \ref{taba3}- \ref{taba5} show the quasinormal frequencies produced by altering the parameters $\rho_s$, $r_s$, and $\gamma$. 
\begin{center}
\begin{tabular}{|c|c|c|}
 \hline 
 \multicolumn{3}{|c|}{ $r_{s}=0.4$, $\gamma=0.02$, $n=0$, $l=2$, $M=1$}
\\ \hline $\rho _{s}$ & $Scalar$ & EM \\ \hline
$0$ & $0.436613-0.085241i$ & $0.415851-0.083537i$ \\ 
$0.2$ & $0.415693-0.08069i$ & $0.39614-0.079091i$ \\ 
$0.4$ & $0.39624-0.076523i$ & $0.37780-0.075007i$ \\ 
$0.6$ & $0.37813-0.072683i$ & $0.36072-0.071248i$ \\ 
$0.8$ & $0.36125-0.069141i$ & $0.34478-0.067779i$ \\ 
$1$ & $0.34550-0.065867i$ & $0.32990-0.06457i$%
\\ 
 \hline
\end{tabular}
\captionof{table}{Variation of amplitude and damping of QNMs with respect to central halo density parameter.} \label{taba3}
\end{center}
\begin{center}
\begin{tabular}{|c|c|c|}
 \hline 
 \multicolumn{3}{|c|}{ $\rho=0.6$, $\gamma=0.02$, $n=0$, $l=2$, $M=1$}
\\ \hline $r_{s}$ & $Scalar$ & EM \\ \hline
$0$ & $0.436613-0.085241i$ & $0.41585-0.083537i$ \\ 
$0.1$ & $0.43545-0.085003i$ & $0.41475-0.083303i$ \\ 
$0.2$ & $0.42787-0.083407i$ & $0.40760-0.081740i$ \\ 
$0.3$ & $0.40935-0.079428i$ & $0.39014-0.077846i$ \\ 
$0.4$ & $0.37813-0.072683i$ & $0.36072-0.071248i$ \\ 
$0.5$ & $0.33561-0.063575i$ & $0.32062-0.062334i$ \\ 
$0.6$ & $0.28566-0.053108i$ & $0.27344-0.052083i$%
\\ 
 \hline
\end{tabular}
\captionof{table}{Variation of amplitude and damping of QNMs with respect to central halo radius parameter.} \label{taba4}
\end{center}
\begin{center}
\begin{tabular}{|c|c|c|}
 \hline 
 \multicolumn{3}{|c|}{ $r_{s}=0.4$, $\rho_s=0.6$, $n=0$, $l=2$, $M=1$}
\\ \hline $\gamma$ & $Scalar$ & EM \\ \hline
$0$ & $0.42563-0.084147i$ & $0.40300-0.082665i$ \\ 
$0.01$ & $0.40227-0.078470i$ & $0.38228-0.077000i$ \\ 
$0.02$ & $0.37813-0.072683i$ & $0.36072-0.071248i$ \\ 
$0.03$ & $0.35307-0.066770i$ & $0.33817-0.065393i$ \\ 
$0.04$ & $0.32690-0.060710i$ & $0.31443-0.059415i$ \\ 
$0.05$ & $0.29938-0.054472i$ & $0.28924-0.053285i$ \\ 
$0.06$ & $0.27011-0.048012i$ & $0.26221-0.046964i$%
\\ 
 \hline
\end{tabular}
\captionof{table}{Variation of amplitude and damping of QNMs with respect to quintessence parameter.} \label{taba5}
\end{center}
The results from Tables \ref{taba3}, \ref{taba4}, and \ref{taba5} are summarised in figures \ref{realRo}, \ref{realS}  and \ref{realG}. It is observed for both perturbations that, increasing the DM halo parameters ($\rho_s$ and $r_{s}$), as well as the quintessence parameter $\gamma$, decreases the magnitudes of both the imaginary and the real parts of the quasinormal frequencies. In general, gravitational waves emitted from BHs surrounded by DM halo and quintessence field will have lower frequency and decay rate compared to those emitted from BHs in vacuum. 

\section{Conclusion} \label{sec6}

In this article, we considered a Schwarzschild BH embedded in a Dehnen-type DM halo with a quintessential field background. To accomplish this, we first derived the metric function and then discussed the impact of the DM and quintessence field components in three different scenarios throughout the rest of the article. The first important attribute of black holes is their singularity and horizons. We discovered that the BH solution's singularity at $r = 0$ is a necessary singularity that no coordinate translation can remove. It is notable that the BH described by the metric (\ref{m1}) has two horizons: the event horizon and the cosmological horizon. Both horizons are mostly dependent on the parameters of the dark sector. It was found that all dark sector parameters $(\rho_s,r_s, \gamma)$ increase $r_h$ while lowering $r_c$. 

We then concentrated on the BH shadow analysis. After determining the effective potential, we estimated the photon and shadow radii. We demonstrated that the photon sphere and shadow radius grow with core density and radius. Similarly, both radii expand in proportion to the quintessence parameter. We hypothesised that the presence of Dehnen type DM and the quintessence field increase the size of the shadow. Additionally, relying on EHT observations of M87$^{\star}$ and Sgr A$^{\star}$, we estimated the best-fit constraints on the DM density parameter $\rho_s$ and the quintessence parameter $\gamma$. Our estimations showed that the DM density $\rho_s$ can have values up to $<0.48$ and $<0.12$ in the quintessence field background with the parameter $\gamma<0.035$ and $<0.01$ for the observational data of M87$^{\star}$ and Sgr A$^{\star}$, respectively.

Further, we delved into the role of the Dehnen-type BH parameters in the presence of a quintessence field on the deflection angle in the weak approximation. This is what we examined to explore the unique aspects of the BH parameters by bringing out their combined effects on the deflection angle around the Dehnen type DM BH surrounded by the quintessence field. We found that the deflection angle changes at a similar rate due to the impacts of the BH parameters $(\rho, r_s, \gamma)$, resulting in the deflection angle shifting upward and increasing to possibly larger values. Furthermore, through analysis, we provided a physical interpretation of the parameters $(\rho, r_s, \gamma)$ with similar effects that cause the deflection angle to increase. This is consistent with the physical interpretation of these parameters acting as repulsive gravitational charges that can weaken the BH background gravity. 

Finally, QNMs are investigated using the 6th order WKB approach. Since we have limited ourselves to the case of $n = 0$ and $l=2$, the 6th order WKB approach yields accurate results. We considered both scalar and electromagnetic perturbations. It is found that, increasing the DM halo parameters ($\rho_s$ and $r_s$), as well as the quintessence parameter $\gamma$, reduces the magnitudes of both the imaginary and real parts of the quasinormal frequencies. In addition, we showed that this BH is stable to perturbations. We conclude that the presence of the DM halo and the quintessence field causes gravitational waves to propagate slower than a  Schwarzschild BH in vacuum. 

In conclusion, our results show that the simultaneous presence of stretch type DM and quintessence field leads to interesting effects on Schwarzschild BH horizons, shadows and deflection angle as well as QNMs.

\section*{ACKNOWLEDGEMENT}

The research is supported by the National Natural Science Foundation of China under Grant No. W2433018.


\begin{thebibliography}{the}


\bibitem{Borde03PRL} A. Borde, A. H. Guth, and A. Vilenkin, Phys. Rev. Lett. 90, 151301 (2003), arXiv:gr-qc/0110012 [gr-qc].

\bibitem{Borde94PRL} A. Borde and A. Vilenkin, Phys. Rev. Lett. 72, 3305 (1994), arXiv:gr-qc/9312022 [gr-qc].

\bibitem{Abbott16a}B. P. Abbott and et al. (Virgo and LIGO Scientific Collaborations), Phys. Rev. Lett. 116,
061102 (2016), arXiv:1602.03837 [gr-qc].

\bibitem{Abbott16b} B. P. Abbott and et al. (Virgo and LIGO Scientific Collaborations), Phys. Rev. Lett. 116,
241102 (2016), arXiv:1602.03840 [gr-qc].

\bibitem{Akiyama19L1}K. Akiyama and et al. (Event Horizon Telescope Collaboration), Astrophys. J. 875, L1 (2019),
arXiv:1906.11238 [astro-ph.GA].

\bibitem{Akiyama19L6}K. Akiyama and et al. (Event Horizon Telescope Collaboration), Astrophys. J. 875, L6 (2019),
arXiv:1906.11243 [astro-ph.GA].
\bibitem{Peebles03} P. J. Peebles and B. Ratra, Reviews of Modern Physics 75, 559 (2003), astro-ph/0207347.

\bibitem{Spergel07} D. N. Spergel, R. Bean, and et al. (WMAP), Astro-phys. J. Suppl. 170, 377 (2007), arXiv:astro-ph/0603449.

\bibitem{Stuchlik05} Z. Stuchlık, Mod. Phys. Lett. A 20, 561 (2005), arXiv:0804.2266. 

\bibitem{Cruz05} N. Cruz, M. Olivares, and J. R. Villanueva, Class. Quantum Grav. 22, 1167 (2005), gr-qc/0408016.

\bibitem{Stuchlik11}  Z. Stuchl´ık and J. Schee, J. Cosmol. Astropart. Phys. 9, 018 (2011).

\bibitem{Grenon10} C. Grenon and K. Lake, Phys. Rev. D 81, 023501 (2010), arXiv:0910.0241. 

\bibitem{Rezzolla03a} L. Rezzolla, O. Zanotti, and J. A. Font, Astron. Astrophys. 412, 603 (2003), gr-qc/0310045.

\bibitem{Arraut15} I. Arraut, Int. J. Mod. Phys. D 24, 1550022 (2015)

\bibitem{Faraoni15} V. Faraoni, ed., Lecture Notes in Physics, Berlin Springer Verlag Vol. 907 (2015).

\bibitem{Shaymatov18a} S. Shaymatov, B. Ahmedov, Z. Stuchlik, and A. Abdujabbarov, 	Int. J. Mod. Phys. D 27, 1850088 (2018).

\bibitem{Kiselev03} V. V. Kiselev, Class. Quantum Grav. 20, 1187 (2003), gr-qc/0210040.

\bibitem{Wetterich88} C. Wetterich, Nucl. Phys. B 302, 668 (1988).

\bibitem{Caldwell09}  R. Caldwell and M. Kamionkowski, Nature 458, 587 (2009). 

\bibitem{Hellerman2001JHEP} S. Hellerman, N. Kaloper, and L. Susskind, J. High Energy Phys. 2001, 003 (2001), arXiv:hep-th/0104180.

\bibitem{Persic96} M. Persic, P. Salucci, and F. Stel, Mon. Not. R. Astron. Soc. 281, 27 (1996), arXiv:astro-
ph/9506004 [astro-ph]. 

\bibitem{Li-Yang12} M.-H. Li and K.-C. Yang, Phys. Rev. D 86, 123015 (2012), arXiv:1204.3178 [astro-ph.CO].


\bibitem{Hendi20} S. H. Hendi, A. Nemati, K. Lin, and M. Jamil, Eur. Phys. J. C 80, 296 (2020),
arXiv:2001.01591 [gr-qc].


\bibitem{Rizwan19} M. Rizwan, M. Jamil, and K. Jusufi, Phys. Rev. D 99, 024050 (2019), arXiv:1812.01331
[gr-qc].

\bibitem{Narzilloev20b} B. Narzilloev, J. Rayimbaev, S. Shaymatov, A. Abdujabbarov, B. Ahmedov, and C. Bambi,
Phys. Rev. D 102, 104062 (2020), arXiv:2011.06148 [gr-qc].

\bibitem{Shaymatov21d} S. Shaymatov, B. Ahmedov, and M. Jamil, Eur. Phys. J. C 81, 588 (2021).

\bibitem{RayimbaevShaymatov21a} J. Rayimbaev, S. Shaymatov, and M. Jamil, Eur. Phys. J. C 81, 699 (2021), arXiv:2107.13436
[gr-qc].

\bibitem{Shaymatov21pdu} S. Shaymatov, D. Malafarina, and B. Ahmedov, Phys. Dark Universe 34, 100891 (2021),
arXiv:2004.06811 [gr-qc].

\bibitem{Shaymatov22prd} S. Shaymatov, P. Sheoran, and S. Siwach, Phys. Rev. D 105, 104059 (2022), arXiv:2110.10610 [gr-qc]. 

\bibitem{Cardoso22DM} V. Cardoso, K. Destounis, F. Duque, R. P. Macedo, A. Maselli, Phys. Rev. D105, L061501 (2022), arXiv:2109.00005 [gr-qc].

\bibitem{Shen24PLB} Z. Shen, A. Wang, Y. Gong, Sh. Yin, Phys. Lett. B 855 138797 (2024), arXiv:2311.12259 [gr-qc]. 

\bibitem{Hou18-dm} X. Hou, Z. Xu, M. Zhou, and J. Wang, J. Cosmol. Astropart. Phys. 2018, 015 (2018), arXiv:1804.08110 [gr-qc].

\bibitem{Dehnen93} W. Dehnen, Mon. Not. R. Astron. Soc. 265, 250 (1993).

\bibitem{dn14} Mrinnoy M. Gohain, Prabwal Phukon and Kalyan Bhuyan. arXiv:2407.02872 (2024).

\bibitem{qqn1} K. Akiyama and et al. Astrophys. J. Lett. 930(2):L12, 2022.

\bibitem{qqn2} K. Akiyama and et al. Astrophys. J. Lett. 930(2):L14, 2022.

\bibitem{qqn3}K. Akiyama and et al. Astrophys. J. Lett. 930(2):L15, 2022.

\bibitem{qqn4}K. Akiyama and et al. Astrophys. J. Lett. 930(2):L17, 2022. 

\bibitem{Synge66} J. L. Synge, Mon. Not. R. Astron. Soc. 131, 463 (1966).

\bibitem{Luminet79} J.-P. Luminet, 	Astron. Astrophys. 75, 228 (1979).

\bibitem{Amarilla13} L. Amarilla and E. F. Eiroa, Phys. Rev. D 87, 044057 (2013).

\bibitem{Konoplya19} R. A. Konoplya, Phys. Lett. B 795, 1 (2019).

\bibitem{Vagnozzi19} S. Vagnozzi and L. Visinelli, Phys. Rev. D 100, 024020 (2019). 


\bibitem{Afrin21a} M. Afrin, R. Kumar, and S. G. Ghosh, Mon. Not. R. Astron. Soc. 504, 5927 (2021), arXiv:2103.11417 [gr-qc].


\bibitem{Atamurotov16EPJC} F. Atamurotov, S. G. Ghosh, and B. Ahmedov, Eur. Phys. J. C 76, 273 (2016), arXiv:1506.03690 [gr-qc].

\bibitem{Konoplya19PRD} R. A. Konoplya and A. Zhidenko, Phys. Rev. D 100, 044015 (2019).

\bibitem{Atamurotov21JCAP} F. Atamurotov, S. Shaymatov, P. Sheoran, and S. Siwach, J. Cosmol. Astropart. Phys. 2021 (8), 045, arXiv:2105.02214 [grqc]. 

\bibitem{Mustafa22CPC} G. Mustafa, F. Atamurotov, I. Hussain, S. Shaymatov, and A. Ovgun, Chin. Phys. C 46, 125107 (2022),
arXiv:2207.07608 [gr-qc].

\bibitem{Tsukamoto18} N. Tsukamoto, Phys. Rev. D 97, 064021 (2018).


\bibitem{Asukula24} H. Asukula, S. Bahamonde, M. Hohmann, V. Karanasou, C. Pfeifer, and J. L. Rosa, Phys. Rev. D 109, 064027 (2024), arXiv:2311.17999 [gr-qc]. 

\bibitem{Rosa23b}  J. L. Rosa, Phys. Rev. D 107, 084048 (2023). 

\bibitem{Moffat20} J. W. Moffat and V. T. Toth, Phys. Rev. D 101, 024014 (2020), arXiv:1904.04142 [gr-qc].

\bibitem{Al-BadawiCTP24} A. Al-Badawi, S. Shaymatov, M. Alloqulov, A. Wang, Commun. Theor. Phys. 76,  085401 (2024).

\bibitem{Al-BadawiCPC24} A. Al-Badawi, M. Alloqulov, S. Shaymatov, B. Ahmedov, Chin. Phys. C 48,  095105 (2024).

\bibitem{Hendi23} S. Hendi, K. Jafarzade, and B. Eslam Panah, J. Cosmol. Astropart. Phys. 2023 (02), 022 (2023).
\bibitem{newa1}Yan, Z., Zhang, X., Wan, M. et al. Shadows and quasinormal modes of a charged non-commutative black hole by different methods. Eur. Phys. J. Plus 138, 377 (2023).

\bibitem{Eddington1919GL} A. S. Eddington, The Observatory 42, 119 (1919).

\bibitem{Bisnovatyi-Kogan2010a} G. S. Bisnovatyi-Kogan and O. Y. Tsupko, Mon. Not. R. Astron. Soc. 404, 1790 (2010).

\bibitem{Tsupko12} O. Y. Tsupko and G. S. Bisnovatyi-Kogan, Gravitation and Cosmology 18, 117 (2012).

\bibitem{Cunha20a} P. V. P. Cunha, N. A. Eiro, C. A. R. Herdeiro, and J. P. S. Lemos, J. Cosmol. A. P 2020, 035 (2020), arXiv:1912.08833 [gr-qc] .

\bibitem{Babar21a} G. Z. Babar, F. Atamurotov, and A. Z. Babar, Phys. Dark Universe 32, 100798 (2021).

\bibitem{Javed22GRG} W. Javed, M. Atique, and A. {\"O}vg{\"u}n, Gen. Relativ. Gravit. 54, 135 (2022), arXiv:2210.17277 [gr-qc].  

\bibitem{Jafarzade21a} K. Jafarzade, M. Kord Zangeneh, and F. S. N. Lobo, J. Cosmol. Astropart. Phys. Phys. 2021, 008 (2021), arXiv:2010.05755 [gr-qc].

\bibitem{Atamurotov22} F. Atamurotov, D. Ortiqboev, A. Abdujabbarov, and G. Mustafa, Eur. Phys. J. C 82, 659 (2022).

\bibitem{Atamurotov21galaxy}  F. Atamurotov, S. Shaymatov, and B. Ahmedov, Galaxies 9, 54 (2021).

\bibitem{Kumaran_2023} Y. Kumaran and A. {\"O}vg{\"u}n, Eur. Phys. J. C 83 812 (2023).

\bibitem{Mizuno_2018} Y. Mizuno, Z. Younsi, C. M. Fromm, O. Porth, M. De Laurentis, H. Olivares, H. Falcke, M. Kramer, and L. Rezzolla, Nature Astronomy 2, 585–590 (2018).

\bibitem{Rahvar19MNRAS} S. Rahvar and J. W. Moffat, MNRAS 482, 4514 (2019),
arXiv:1807.07424 [gr-qc].

\bibitem{Izmailov19MNRAS} R. N. Izmailov, R. K. Karimov, E. R. Zhdanov, and K. K. Nandi, MNRAS 483, 3754 (2019), arXiv:1905.01900 [gr-qc]. 


\bibitem{newa2}Chen Wu,  Int.J.Mod.Phys.D 26 (2017) 10, 1750111.

\bibitem{newa3}Zening Yan, Chen Wu, Wenjun Guo, Nucl.Phys.B 973 (2021) 115595.

\bibitem{d1} H. Mo, F. van den Bosch, and S. White, Galaxy Formation and Evolution (Cambridge University Press, Cambridge,
England, UK, 2010)

\bibitem{ds1}J. R. Shakeshaft, The Formation and Dynamics of Galaxies (Dordrecht, The Netherlands, 2012).

\bibitem{dn9} Z. Xu, X. Hou, X. Gong, et al., J. Cosmol. Astropart. Phys. 2018, 038 (2018).

\bibitem{dn10} M. Azreg-A\"inou, Phys. Rev. D - Part. Fields, Gravit. Cosmol. 90 (2014).

\bibitem{dn11} K. Jusufi, J. Mubasher, and Z. Tao, Eur. Phys. J. C. 80, 354 (2020).

\bibitem{dn12} X. Hou, Z. Xu, M. Zhou, et al., J. Cosmol. Astropart. Phys. 2018, 015 (2018).

\bibitem{dn13}Z. Xu, J. Wang, and M. Tang, J. Cosmol. Astropart. Phys. 2021, 007 (2021).




\bibitem{Gibbons08CQG} G. W. Gibbons and M. C. Werner, Class. Quant. Grav. \textbf{25}, 235009 (2008).

\bibitem{Werner12GBT} M. C. Werner, Gen. Relativ. Gravit. \textbf{44}, 3047 (2012).

\bibitem{Ono17GBT} T. Ono, A. Ishihara, and H. Asada,  Phys. Rev. D \textbf{96}, 104037 (2017).

\bibitem{Ishihara16GBT}  A. Ishihara et al, Phys. Rev. D \textbf{94}, 084015 (2016).

\bibitem{Ishihara16} A. Ishihara, Y. Suzuki, T. Ono, and H. Asada, Phys. Rev. D \textbf{95}, 044017 (2017). 


\bibitem{Jusufi18GBT} K. Jusufi, A.~Ovgun, J. Saavedra, Y. Vasquez, and P. A. Gonzalez, Phys. Rev. D \textbf{97}, 124024 (2018).

\bibitem{Li20} Z. Li and A. Ovg\"un, Phys. Rev. D  \textbf{101}, 024040 (2020).

\bibitem{Zhang21GBT} Z. Zhang, Class. Quant. Grav. \textbf{39}, 015003 (2022).

\bibitem{DCarvalho21} I. D. D. Carvalho, G. Alencar, W. M. Mendes, and R. R. Landim, EPL \textbf{134}, 51001 (2021).

\bibitem{isMandal:2023eae} S.~Mandal, Phys. Dark Univ. \textbf{42}, 101374 (2023). 

\bibitem{Al-Badawi24EPJC_GBTa} A. Albadawi, S. Shaymatov, and \.{I}.~Sakall\i{}, Eur. Phys. J. C \textbf{84}, 825 (2024).

\bibitem{Al-Badawi24EPJC_GBTb} A. Albadawi, S. Shaymatov, S. K. Jha, and A. Rahaman, Eur. Phys. J. C \textbf{84}, 722 (2024).

\bibitem{Wald:1984} R. M. Wald, \textit{General Relativity} (Chicago Univ. Pr., Chicago, 1984).

\bibitem{Mandal:2023} S. Mandal, Phys. Dark Univ. \textbf{42}, 101374 (2023).

\bibitem{Chandrasekhar:1985} S. Chandrasekhar, \textit{The Mathematical Theory of Black Holes} (Chicago Univ. Pr., Chicago, 1985).

\bibitem{Sakalli:2016} I. Sakalli and A. Ovgun, J. Astrophys. Astron. \textbf{37}, 21 (2016).



\bibitem{d3} S. Iyer, C.M. Will, Phys. Rev. D 35, 3621 (1987).

\bibitem{d4} R.A. Konoplya, Phys. Rev. D 68 024018 (2003).


\end{thebibliography}
\end{document}